\documentclass[12pt]{article}
\usepackage{amsfonts,amsthm,amsmath,amssymb,upgreek,bm}
\usepackage{hyperref}
\usepackage[paper=letterpaper,margin=.85in]{geometry}
\usepackage{graphicx}
\usepackage{subfigure}
\usepackage{units}
\usepackage{float}
\usepackage{svg}
\usepackage[export]{adjustbox}
\usepackage{xcolor}
\usepackage[shortlabels]{enumitem}
\usepackage[mathscr]{euscript}
\usepackage[normalem]{ulem}
\usepackage{bm}
\usepackage{comment}

\newcommand{\be}{\begin{equation}}
\newcommand{\ee}{\end{equation}}
\newcommand{\bea}{\begin{eqnarray}}
\newcommand{\eea}{\end{eqnarray}}
\renewcommand{\tilde}{\widetilde}
\renewcommand{\i}{\mathrm{i}}
\renewcommand{\d}{\mathrm{d}}

\numberwithin{equation}{section}

\def\Tr{\text{Tr}}

\usepackage{eso-pic}% http://ctan.org/pkg/eso-pic
\usepackage{lipsum}% http://ctan.org/pkg/lipsum

\begin{document}
\thispagestyle{empty}
\AddToShipoutPictureBG*{%
  \AtPageUpperLeft{%
    \hspace{\paperwidth}%
    \raisebox{-\baselineskip}{%
      \makebox[0pt][r]{USTC-ICTS/PCFT-25-58~~~}
}}}%

\vspace*{2.5cm}
\begin{center}

{\bf {\LARGE Comments on the de Sitter Double Cone}}\\

\begin{center}

\vspace{1cm}

{\bf Zhenbin Yang$^{1,2}$,  Yuzhen Zhang$^{3}$ and Wenwen Zheng$^{3,4}$}\\
 \bigskip \rm

\bigskip 

${}^1$Institute for Advanced Study, Tsinghua University\\

${}^2$Peng Huanwu Center for Fundamental Theory, Hefei, Anhui 230026, China

${}^2$Department of Physics, University of California, Santa Barbara

${}^3$School of Physics, Zhejiang University

\rm
  \end{center}

\vspace{2.5cm}
{\bf Abstract}
\end{center}
\begin{quotation}
\noindent

We study the double cone geometry proposed by Saad, Shenker, and Stanford in de Sitter space. We demonstrate that with the inclusion of static patch observers, the double cone leads to a linear ramp consistent with random matrix behavior. This ramp arises from the relative time shift between two clocks located in opposite static patches.
\end{quotation}

\setcounter{page}{0}
\setcounter{tocdepth}{2}
\setcounter{footnote}{0}
\newpage

\parskip 0.1in
 
\setcounter{page}{2}
\tableofcontents

\newpage

\section{Introduction}

Does time exist forever in our universe? 

Or, to phrase it differently, is there a perfect clock in de Sitter space that can function eternally? Let's consider the situation that a perfect clock runs for a duration $\mathcal{T}$ in de Sitter space dS$_D$. Being a perfect clock means the state $|\mathcal{T}\rangle$ of the clock at different times should be orthogonal, namely the return amplitude should be zero:
\be\label{eqn:clockra}
\Tr_{\text{clock}} e^{\i \mathcal{E} \mathcal{T}}=\int \d \mathcal{T}_0\langle \mathcal{T}|0\rangle=\delta(\mathcal{T})\times \int \d \mathcal{T}_0,
\ee
where $\mathcal{E}$ is defined later to be the energy of the clock.
Nevertheless, when we put the clock into the de Sitter space, then the finite entropy $S_{\text{dS}}$ of the de Sitter horizon suggests that such behavior cannot persist forever\cite{Dyson:2002pf,Goheer:2002vf}, due to the discreteness of the spectrum. While the existence of a precise microscopic description for the de Sitter static patch remains an open question, it is worthwhile to ask whether general relativity in de Sitter space itself inherently prevents the existence of an ideal clock.

\begin{figure}\label{fig:DC}
\centering
\includegraphics[width=8cm]{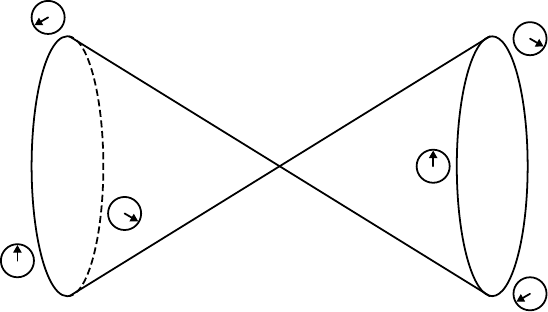}
\caption{A de Sitter double cone with the presence of two clocks.}
\end{figure}

In this letter, we argue that such a mechanism arises naturally within de Sitter geometry. Specifically, we generalize the double cone geometry \cite{Saad:2018bqo,Chen:2023hra}, originally introduced by Saad, Shenker, and Stanford in the context of black hole systems, to de Sitter space. By considering two clocks evolving in opposite directions, placed on opposite sides of the de Sitter static patch, we construct a quotient of de Sitter space with respect to a suitably modified boost Killing vector (see Fig.~\ref{fig:DC}). The resulting geometry introduces nonperturbative corrections (of order $e^{-2 S_{\text{dS}}}$) to the return probability of the system over the clock evolution time $\mathcal{T}$ before the Heisenberg time $e^{ S_{\text{dS}}}$.

From a high-level perspective, the double cone geometry can be viewed as a certain modified trace over the semiclassical bulk Hilbert space. In the black hole case, the bulk Hilbert space is $\mathcal{H}_{\text{QFT}}\otimes L^2(\mathbb{R})$ \cite{Witten:2021unn,Chandrasekaran:2022eqq} with the later part describes the relative time shift between the thermofield double state, and the double cone partition function calculates:
\be\label{eqn:Zdcintro}
Z_{\text{DC}}=\Tr_{L^2(\mathbb{S}^1)}[f(m)^2] \Tr_{\text{QFT}} e^{-\i \tilde K_m T}.
\ee
with $m$ represents the (shifted) mass of the black hole and $L^2(\mathbb{S}^1)$ is a notation used in \cite{Chen:2023hra} to represent a by-hand compactification of the relative time shift to period $T$ during the evaluate of $\Tr_{L^2(\mathbb{R})}$. The $\tilde K_m$ is a modified boost operator acting on $\mathcal{H}_{\text{QFT}}$ defined by the double cone geometry \eqref{eqn:QFTtr}, whose spectrum is given by the quasi-normal frequencies of the system. The $f(m)$ factor is a window function introduced to constrain the energy integral that will be taken to be a gaussian distribution. It is dual to the same factor inserted when we consider the microcanonical Spectral Form factor (SFF):\footnote{We use the convention that the subscribe $f$ means insertion of window function $f(\mathcal{E})$ in the partition function.}
\be\label{eqn:SFF}
|Z_f(\i T)|^2=|\Tr_{\text{bdy}}(f(H)e^{-\i T H})|^2.
\ee

In the de Sitter case, we essentially have the same expression. It is already shown in \cite{Chandrasekaran:2022cip} that the semiclassical bulk Hilbert space of de Sitter in the presence of clocks in the two static patch takes the form $\tilde{\mathcal{H}}_{\text{QFT}}\otimes L^2(\mathbb{R}_+)$, where the first part is the bulk QFT Hilbert (with the constraint of rotation invariance see \eqref{eqn:Hbulk} ) and the later part describes the relative time shift between the clocks at the two sides. Effectively, the clock plays the role of the gravitational boundary and we will show in \eqref{eqn:ZDC} that the double cone evaluates the same bulk trace as in equation \eqref{eqn:Zdcintro}, where $m$ now is replaced by the energy of the clock $\mathcal{E}.$ The purpose of the letter is to fill in technical details of the derivation and also try to give a putative holographic interpretation analogous to \eqref{eqn:SFF}.

We will analyze this problem using the gravitational path integral for de Sitter space with the presence of an observer. The role of observers in de Sitter space has been a central theme in prior studies \cite{Dong:2018cuv, Chandrasekaran:2022cip,Silverstein:2022dfj,Kaplan:2024xyk,Kolchmeyer:2024fly,Tietto:2025oxn}, and more recently in \cite{Maldacena:2024spf},\footnote{See also \cite{Ivo:2025yek,Shi:2025amq}.} which demonstrated that, up to an important minus sign that will be mentioned later, the inclusion of an observer modified the sphere path integral to be consistent with a state counting interpretation. We follow closely the observer model studied in \cite{Maldacena:2024spf}, which consists of a relativistic particle of large mass (\(m \gg 1\)) to keep track of the spacetime location of the observer and an internal clock with energy \(\mathcal{E}\) that measures the proper time of the observer. In \cite{Maldacena:2024spf}, it was argued that the de Sitter sphere partition function conditioned on the existence of an observer supports a quantum mechanical interpretation in terms of a union of the observer's Hilbert space and the de Sitter Hilbert space subject to the Hamiltonian constraint. Here the observer's Hilbert space consists of both the particle and the clock:
\be\label{eqn:Hilbertstru}
\mathcal{H}_{\mathrm{obs}} = \mathcal{H}_{\mathrm{particle}} \otimes \mathcal{H}_{\mathrm{clock}},
\ee
and the de Sitter Hilbert space associated with the de Sitter horizon, \(\mathcal{H}_{\text{dS}}\), describes the rest of the universe. Since the universe is closed, the total Hilbert space is subject to the Hamiltonian constraint:
\be\label{eqn:Hconint}
\mathbf{H} = H_{\text{dS}} + H_{\mathrm{obs}} = 0, \quad 
\mathcal{H} = \frac{\mathcal{H}_{\text{dS}} \otimes \mathcal{H}_{\mathrm{obs}}}{\mathbf{H} = 0}.
\ee
\cite{Maldacena:2024spf} showed that the sphere partition function $Z_{\text{sphere}}$ can be decomposed as:\footnote{Throughout the paper, $\sim$ means asymptotes at large $\mathcal{T}$ or large $m$.}
\be\label{eqn:Zsphere}
Z_{\text{sphere}} =  \int {\mathrm{d} \beta\over \beta} \, Z_{\mathrm{patch}}(\beta) Z_{\mathrm{particle}}(\beta) Z_{\mathrm{clock}}(\beta) \equiv -\i \Tr_{\mathcal{H}}\mathbf{1}\sim -\i e^{S_{\text{dS}}} {e^{-2\pi m} m^{D-1}\over 2\pi (D-1)!}e^{S_{\text{clock}}}.
\ee 
Here, compared with \cite{Maldacena:2024spf}, we have included a ${1\over \beta}$ measure factor in the $\beta$ integral that represents the quotient of the U(1) symmetry associated with time shift. Correspondingly, \(Z_{\mathrm{patch}}(\beta)\) represents the sphere gravitational partition function  with a fixed Euclidean geodesic circle length at the particle's location with a marked point, that is differ from \cite{Maldacena:2024spf} by a factor of $\beta$. All the rest is the same as \cite{Maldacena:2024spf}, \(Z_{\mathrm{particle}}(\beta)\) and \(Z_{\mathrm{clock}}(\beta)\) represent the partition functions of the particle winding around the circle and the clock, respectively. The integral is analyzed through saddle point approximation at $\beta=2\pi$. The final expression, \( \Tr_{\mathcal{H}}\mathbf{1}\)\footnote{It is denoted as $-Z_{\text{count}}$ in \cite{Maldacena:2024spf}.}, counts the total number of states in the Hilbert space \(\mathcal{H}\), which grows exponentially with de Sitter entropy as\footnote{The $2\pi$ factor arises because the time shift U(1) symmetry in \(Z_{\mathrm{patch}}(\beta)\) has not been quotiented out.}
\be
e^{S_{\text{dS}}}=2\pi e^{A\over 4G_N}|Z_{S^D}|,
\ee 
including one-loop corrections from graviton and matter $|Z_{S^D}|$. The ${e^{-2\pi m} m^{D-1}\over (D-1)!}$ factor comes from the one-loop integral of the particle, and $e^{S_{\text{clock}}}=\int{ \d \mathcal{T}_0\over 2\pi}$ represents the zero mode integral of the clock as in \eqref{eqn:clockra}.
In this letter, we will be interested in the path integral with the presence of a marked point, this amounts to remove the ${1\over \beta}$ factor in \eqref{eqn:Zsphere}, that modifies the sphere partition function into:
\be\label{eqn:Zsphere'}
Z_{\text{sphere}}' =  \int \mathrm{d} \beta \, Z_{\mathrm{patch}}(\beta) Z_{\mathrm{particle}}(\beta) Z_{\mathrm{clock}}(\beta)=-2\pi \i \Tr_{\mathcal{H}}\mathbf{1}.
\ee
\cite{Maldacena:2024spf} argued that the factor of $\i$ can be understood as the difference between the measure used in Euclidean path integral and the inverse Laplace transformation of $\beta\rightarrow \beta_0+\i s$ to impose the Hamiltonian constraint. However, the overall minus sign remains unexplained. 
{\bf Note added in V2: } A recent proposal to address the issue of this minus sign via a different prescription of implementing the Hamiltonian constraint is discussed in \cite{Chen:2025jqm}.
% In this letter, we do not address this minus sign and focus instead on the double cone partition function, which we argue to be positive definite. This comes from two facts, the first is that the $\Tr_{L^2(\mathbb{S}^1)}[f(m)^2]$ part in \eqref{eqn:Zdcintro} is manifestly positive, the second fact is that the quasi-normal frequencies always have positive definite imaginary part with both positive and negative real parts (when the real part exists). This means the $\Tr_{\text{QFT}} e^{-\i \tilde K T}$ term is a positive decaying function, and at late times, it asymptotes to a constant value of one, given by the Hartle-Hawking state. Nevertheless, we emphasize that our microscopic interpretation in \eqref{eqn:Zint} crucially relies on the minus sign {\it not} invalidating the state-counting interpretation of the sphere partition function.

Once the clock is inside the de Sitter space, the trace in \eqref{eqn:clockra} is taken over the whole Hilbert space $\mathcal{H}$, and we define the return amplitude as the path integral over geometries containing an observer loop whose clock time evolves from \(0\) to \( \mathcal{T} \):  
\be\label{eqn:ZTdef}
Z(i\mathcal{T}) := \sum_{\text{geometries with observer}} e^{-I_{\text{EH}} - I_{\text{obs}}} \Big|_{0}^{ \mathcal{T}}.
\ee
Here, a slight difference with \eqref{eqn:clockra} is that we have fixed the initial time of the clock to be $0$, which amounts to quotient out the zero mode $\mathcal{T}_0$. This is adopted for convenience, avoiding overall factors of $e^{S_{\text{clock}}}$ in the final expressions.
Sometimes, we will also refer to \( Z(i\mathcal{T}) \) as the Lorentzian partition function.
Here, $I_{\text{EH}}$ is the standard Einstein-Hilbert action with positive cosmological constant (after setting the de Sitter radius to be one):
\be
-I_{\text{EH}}={1\over 16\pi G_N} \int R-(D-1)(D-2).
\ee
And the observer’s action consists of a relativistic particle action and a free clock that measures the particle’s proper time:  
\be\label{eqn:obsac}
-I_{\text{obs}} = -i m \int_0^{T} \mathrm{d}s 
+ i \int_0^{T} \frac{\mathrm{d}\mathcal{T}(s)}{\mathrm{d}s} \mathcal{E}(s) \,\mathrm{d}s 
- i \int_0^{T} \mathcal{E}(s) \,\mathrm{d}s.
\ee
The first term is the standard relativistic particle action, where \( \mathrm{d}s \) represents the proper time of the particle, and \( T \) is its total proper time.  The second and third terms describe the dynamics of the clock, which plays a central role in this framework: the clock time \( \mathcal{T}(s) \) measures the proper time of the particle through its equation of motion, while \( \mathcal{E}(s) \) represents the clock’s energy.  The objective of this letter is to explicitly evaluate this Lorentzian partition function $Z(i\mathcal{T})$ and its absolute value squared that corresponding to the return probability. Below, we will discuss the underlying intuition informally.

The path integral of the clock system effectively imposes the constraint that the total proper time of the particle must be \(\mathcal{T} \). So the path integral of the relativistic particle reduces to a summation over all closed timelike paths with periodicity \(\mathcal{T} \), with a marked point specified by the location of $0$ of the clock variable along the path. This minor modification has an interesting consequence.
It is well known that the close-loop path integral of a relativistic particle does not naturally compute a trace. Instead, it gives rise to a determinant, that can be seen using the Schwinger parameter \( \tau \), under which the loop diagram takes the form
\be\label{eqn:schwinger}
\int_0^{\infty} \frac{\d \tau}{\tau} \Tr\, e^{\tau \nabla^2} = -\Tr \log \left(-\nabla^2\right).
\ee
Here, the factor of \( 1/\tau \) arises from the volume of the gauge group associated with shift of the parametrization of the loop \cite{Polyakov:1987ez}. By fixing the total proper time to be $\mathcal{T}$ and marking a point on the worldline, we remove this gauge symmetry and turn the path integral into a conventional trace of a non-relativistic particle:
\be
\Tr e^{\mathcal{T} \nabla^2},
\ee
with the interpretation of tracing over the evolution operator of a fixed interval of proper time $\mathcal{T}$.
When we consider the path integral on the double cone geometry, the relative time shift between the two marked points associated with the two observers will lead to a linear \( \mathcal{T} \) growth, representing the ramp behavior of the SFF.  

The delta function constraint of the proper time can be easily seen from the constant mode integral of the energy $\mathcal{E}_0$, which is treated as unbounded that ranges from $-\infty$ to $+\infty$. A more realistic clock model can be obtained by introducing an energy filter, such as a Gaussian distribution $f(\mathcal{E}_0)$ introduced in \eqref{eqn:SFF}, to constrain the energy within an order-one window $\sigma$ around a reference energy $\mathcal{E}_*$. This modification smooths the delta function into a Gaussian distribution while leaving the overall discussion unchanged:
\be\label{eqn:smear}
\delta(T-\mathcal{T})\rightarrow\int  \d \mathcal{E}_0 e^{\i \mathcal{E}_0 (\mathcal{T}-T)}f(\mathcal{E}_0)\equiv {\sigma\over \sqrt{2\pi}}\int \d \mathcal{E}_0 e^{\i \mathcal{E}_0 (\mathcal{T}-T)-{(\mathcal{E}_0-\mathcal{E}_*)^2\over 2\sigma^2}}=e^{\i \mathcal{E}_* (\mathcal{T}-T) -{1\over 2}\sigma^2(\mathcal{T}-T)^2}.
\ee

Throughout this work, we will treat $\mathcal{T}$ as a freely varying parameter, allowing it to become arbitrarily large and examining the possible gravitational corrections that arise in this regime. As we said, this is an idealization in which we introduce a perfect clock with infinite entropy into de Sitter space, effectively acting as an ``external boundary." 
A useful perspective on our idealization is that we are working within a time regime where the static patch of de Sitter space admits a well-defined quantum mechanical description, including a reliable notion of a clock. But over what timescale can such a clock remain well-defined? 

One concern is the quantum fluctuations of the particle's trajectory, which could cause it to dissipate into the de Sitter horizon. Naively, this would suggest that the clock description is only valid over a Hubble timescale. However, in Section \ref{sec:Obs}, we will argue that this issue can always be addressed by using de Sitter isometries to gauge-fix the clock at the center of the static patch. Another way to view this is that in a closed universe, we are ``dressing" the rest of the world relative to the observer. In other words, we can regard the empty de Sitter space as an ambient universe, while the physical universe is defined relative to the observer's location. In this sense, the observer plays a role analogous to that of the ``distant star" in the Mach Principle. This abandon of boundary condition is reminiscent of the statement by Hawking and Ellis \cite{hawking2023large}: 
\begin{quote}
    ``We shall take the local physical laws that have been experimentally determined and shall see what these laws imply about the large-scale structure of the universe." 
\end{quote}
We refer to this concept as the \textit{Hawking-Ellis principle} \cite{Dong:2020uxp}. A concrete example of this arises in JT gravity \cite{Jackiw:1984je,Teitelboim:1983ux}: in the AdS case, the Schwarzian boundary \cite{Jensen:2016pah,Maldacena:2016upp,Engelsoy:2016xyb} evolves in the ambient AdS space, which plays the role of the ``distant star." In the de Sitter case \cite{Svesko:2022txo,Rahman:2022jsf}, where JT gravity arises from the dimensional reduction of dS$_3$, the natural boundary condition is given by the data of the observer which fixes both the value of the dilaton and its normal derivative. This leads to a pair of relativistic particles moving in the ambient dS$_2$ space, that take on the role of the observer \cite{HuYangZhangZheng:upcoming} to dress other excitations.

A second concern about the validity of the clock comes from interactions between the clock and thermal fluctuations. The minimal interaction strength, mediated by graviton exchange, is of order \( G_N \). This perturbation alters the observer’s trajectory and leads to an order-one effect after a scrambling time of \( \log(1/G_N) \) so that the observer itself again goes behind the de Sitter horizon. Such interaction can be thought of as order $G_N$ corrections to the Hamiltonian constraint \eqref{eqn:Hconint}. While we do not have much to say about these interactions in general, it is noteworthy that in lower-dimensional models such as \( dS_2 \) and \( dS_3 \), such effects can be avoided due to the absence of local graviton excitations, allowing the clock to persist longer. Perhaps in higher dimensional case, one can view such thermal fluctuations as fluctuations of the location of the de Sitter horizon instead of the thermalization of the observer.
Furthermore, we will argue that the double cone geometry we construct will be stable after a few Hubble times, after which the matter field will dissipate into the de Sitter horizon, and thus contributes meaningfully even before the scrambling time—albeit with an exponentially small amplitude.

At even longer timescales, nonperturbative effects can become significant, potentially influencing the lifetime of the clock over times of order \( e^{1/G_N} \). The double cone geometry we describe can be viewed as one such nonperturbative effect.\footnote{The de Sitter vacuum may also decay, but for our current purposes we assume an eternal de Sitter space.}  We can view its effect as an intrinsic quantum gravity feature that prevents the existence of a perfect clock in de Sitter space.\footnote{In same spirit as the situation of typical state in black holes \cite{Almheiri:2013hfa,Marolf:2013dba,Susskind:2012rm,Susskind:2015toa,deBoer:2018ibj,DeBoer:2019yoe,Susskind:2020wwe,Stanford:2022fdt,Blommaert:2024ftn,Iliesiu:2024cnh}.}

Let's return to the discussion of the Lorentzian partition function. We have already argued that the clock partition function simply restricts the path integral of the relativistic particle into that of a non-relativistic particle. In a closed universe, we will also need to gauge out the de Sitter isometry, and this will be the subject of Section~\ref{sec:Obs}. There, we will show that after gauge out the de Sitter isometry, the two-sided bulk Hilbert space of the observer will become $L^2(\mathbb{R}_+)$. This agrees with the result of \cite{Chandrasekaran:2022cip} under the assumption of no transverse fluctuations of the observer.
In equation \eqref{eqn:Zltime_f}, we will show that the Lorentzian partition function is given by an insertion of $e^{\i \mathcal{E} \mathcal{T}}$ in the sphere partition function \eqref{eqn:Zsphere'}\footnote{After diving $2\pi e^{S_{\text{clock}}} $ that represents gauging the zero mode of the clock.}. Using the notation in \cite{Maldacena:2024spf}, it is given by\footnote{We use the convention that $\rho_{\text{Patch}}$ is positive.}:
\be\label{eqn:Zf}
\i Z_f(\i \mathcal{T}) \sim { m^{D-1}\over (D-1)!}\int_0^{\infty}  \d \mathcal{E} e^{\i \mathcal{T} \mathcal{E}} \rho_{\text{Patch}}(-m-\mathcal{E}) f(\mathcal{E})\sim \frac{ m^{D-1}}{(D-1)!} e^{S_{\text{dS}} - 2\pi(m + \mathcal{E}_*) + i\mathcal{T}\mathcal{E}_* - \frac{\sigma^2\mathcal{T}^2}{2}} \approx 0.
\ee
This tells us that the leading sphere partition function exhibits exponential decay and goes to zero after time of order $\sqrt{S_{\text{dS}}}$. Such an exponential decay is not surprising in general, as the saddle point approximation could only provide a smooth density of states.

The double cone geometry, on the other hand, arises when evaluating the absolute value squared of the Lorentzian partition function. With a slight abuse of terminology, we will refer to this quantity also as the Spectral Form Factor :
\be\label{eqn:dSSFF}
\text{SFF}\equiv|Z(\i \mathcal{T})|^2 = \sum_{\text{geometries with two observers}} e^{-I_{\text{EH}} - I_{\text{obs,L}} - I_{\text{obs,R}}} \Big|_{\Delta \mathcal{T}_R = \mathcal{T}, \Delta \mathcal{T}_L = -\mathcal{T}}.
\ee
Here, the summation runs over geometries (whether connected or disconnected) that contain two observers, whose clocks wind in opposite directions from $0$ to $\mathcal{T}$ and $0$ to $-\mathcal{T}$. We term this path integral calculating the SFF because its behavior closely mirrors that of the conventional SFF. In the context of AdS space, if the observers are positioned at the asymptotic boundary, this path integral faithfully computes the SFF of the dual quantum mechanical system. 
For the de Sitter case, if we use the microscopic interpretation suggested in \cite{Maldacena:2024spf}, then the clock energy inherits the spectral statistics of $H_{\text{dS}}$ through the Hamiltonian constraint \eqref{eqn:Hconint}. Consequently, the path integral reduces to the SFF for $H_{\text{dS}}$, in parallel with the black hole scenario. In other words, we can view $Z(\i \mathcal{T})$ and its absolute value squared as evaluating the following microscopic quantities (using the relation $H_{\text{obs}}\sim m+\mathcal{E}$):
\be\label{eqn:Zint}
\i Z(\i \mathcal{T})={\Tr_{\mathcal{H}} e^{\i \mathcal{E}\mathcal{T}}\over e^{S_{\text{clock}}}}=\Tr_{\mathcal{H}_{\text{dS}}} e^{-\i H_{\text{dS}}\mathcal{T}-\i m \mathcal{T}},~~~|Z(\i \mathcal{T})|^2=|\Tr_{\mathcal{H}_{\text{dS}}} e^{-\i H_{\text{dS}}\mathcal{T}}|^2.
\ee

On the double cone, the path integral for each clock is identical to the previous analysis, leading to a constraint on two closed timelike geodesics with opposite periodicities, \( \pm \mathcal{T} \) and the integration of the two marked points lead to an overall factor of \( \mathcal{T}^2 \). In the double cone geometry, the two closed timelike geodesics are located at the pode and antipode of the de Sitter space, both are generated by the boost Killing vector. The associated \( U(1) \) isometry of the double cone cancels one factor of \( \mathcal{T} \), resulting in a linear \(\mathcal{T} \) contribution from the clock path integral. Explicitly, we will show that the double cone partition function is given by \eqref{eqn:Zdc}:
\be
|Z_f(\i \mathcal{T})|^2 \supset Z_{\text{DC}}={\mathcal{T} \over 2\pi}\int \d \mathcal{E} \, f(\mathcal{E})^2.
\ee
with $\mathcal{E}$ the average energy of the two clocks.

The remaining part of the path integral involves the evaluation of the matter partition function on the double cone, including contributions from the two particles. In Section \ref{sec:doublecone}, we will argue that after a few Hubble times, the one-loop contribution on the double cone simply becomes equal to one. This arises from two effects: First, for ordinary matter excitations on the double cone, the one-loop integral computes the trace of the evolution operator associated with the modified boost operator, whose eigenvalues are given by the quasi-normal frequencies \cite{Chen:2023hra}. At late times, such evolution projects onto the Hartle-Hawking state, and the trace reduces to one (see Section \ref{sec:dcm}). Second, for the two particles associated to the observers, their trajectories can be gauge-fixed at the pode and antipode of the static patch, and we will argue in Section \ref{sec:obsdc} that the path integral of the particles do not display any quasi-normal decay and therefore is a constant.

It is worth noting that, naively, one might expect an additional zero mode, as we can consider a family of double cone geometries from any Schwarzschild-de Sitter geometry. However, the absence of other matter excitations due to quasi-normal decay (including black holes) shows only the vacuum de Sitter double cone contributes at late times. Another way to see it is that in Schwarzschild-de Sitter geometry, there are no enough isometries to fix the location of the observer.

So what is the meaning of the de Sitter double cone?

Here we will offer two possible perspectives. First, in analogy to the black hole case, the de Sitter double cone describes the underlying level statistics of the Hamiltonian (if it exists) of the de Sitter horizon, as in \eqref{eqn:Zint}, which gives a linear ramp growth in the spectral form factor. From this perspective, our result should not come as a complete surprise, as the double cone construction depends only on the local near-horizon Rindler geometry—a feature universal to both cosmic and black hole horizons. The preceding discussion serves primarily to verify that additional factors, such as relocating the observer from the asymptotic boundary to the interior of the static patch, do not disrupt the construction. On the other hand, this is not entirely trivial, as  we have to argue carefully that the observer itself does not decay in the double cone geometry. Another potential concern is that the de Sitter horizon exhibits different scrambling behavior compared to the black hole case. One might reasonably worry that such scrambling effects could manifest in the construction, as we indeed anticipate for more complicated wormholes \cite{Yang:upcoming}. Fortunately, as far as we can tell, the double cone geometry does not explicitly depend on shockwave scattering, allowing a robust construction for any killing horizons including the cosmic horizon.
It is worth emphasizing that from this point of view the emergence of the ramp behavior hinges critically on the inclusion of the clock system. Without it, there is no physical distinction between the various twisted double cone geometries, and consequently, no ramp arises.\footnote{Note added in V2: After published the first version, we learned similar ideas have been thought independently by \cite{ShaghoulianTalks}.} This observation aligns with the broader expectation that the de Sitter static patch, in the absence of an observer, may lack a well-defined quantum mechanical description. After all, how can one meaningfully define a static patch without reference to any physical bodies?\footnote{That said, \cite{Dyson:2002pf,Goheer:2002vf,Susskind:2021omt} has proposed that a statistical description could still be viable in such scenarios, providing a potential pathway to address these conceptual challenges.}

An alternative perspective is related to the question asked at the beginning, that there may not exist a perfect clock in de Sitter space.\footnote{We thank Douglas Stanford for emphasizing this interpretation.} We start with the assumption of a continuously flowing clock with infinite entropy that never returns to its initial state in de Sitter space. However, the double cone wormhole implies that there is always a small, non-perturbative probability of the clock reverting to a previous state. This suggests that the clock’s states will inevitably become random on a timescale of  $e^{S_{\text{dS}}}$, breaking the concept of an eternally progressing clock. 
From this perspective, this problem seems to be related to the typical state firewall paradox of black hole systems, where for a typical state, the notion of time ceases to exist. 
In the current case, one can ask if no perfect clock exists in our universe, what becomes of time?

The outline of the letter is the following:  
\begin{itemize}
    \item In Section~\ref{sec:Obs}, we study the canonical quantization of the observer system in de Sitter space. Using the ultralocal measure, we first demonstrate that prior to gauging out the de Sitter isometries, incorporating the clock degree of freedom maps the observer model to a one-dimensional nonlinear sigma model with target space dS$_D$, where the Hamiltonian corresponds to the clock energy $\mathcal{E}$. After gauge-fixing the de Sitter isometries via the coinvariant formalism, the Hilbert space reduces to $L^2(\mathbb{R}_+)$. We then solve its eigensystem and employ these results to compute the sphere contribution to the Lorentzian partition function.
    \item In Section~\ref{sec:doublecone}, we investigate the de Sitter double cone geometry. We begin by reviewing the general construction of double cone geometries following \cite{Saad:2018bqo}, along with the ordinary matter partition function on these geometries \cite{Chen:2023hra}. Our primary focus is the analysis of the observer's partition function on the double cone. The key conceptual challenge lies in understanding the quasi-normal mode decay of the observer, which we propose is precisely canceled by a measure factor $\mu(T)$ associated with the double cones (see \eqref{eqn:mut}). The physical justification for this measure factor stems from the coinvariant Hilbert space construction, where the observer's evolution manifestly preserves unitarity and thus cannot decay.
    \item This work employs multiple time variables, distinguished as follows:
\begin{itemize}
    \item $\mathcal{T}$: Lorentzian period of the clock
    \item $T$: Bulk Lorentzian period 
    \item $t$: Bulk Lorentzian time
    \item $u$: Observer's Lorentzian proper time (identified with Lorentzian clock time)
    \item $\tau$: Observer's Euclidean proper time (identified with Euclidean clock time)
    \item $\mathbf{T}$: Euclidean clock time prior being identified to the observer's proper time
    \item $s$: Observer's proper time prior being identified to the clock time
\end{itemize}
\end{itemize}

\section{Observer}\label{sec:Obs}
Let's begin by discussing the path integral of the observer. As mentioned in the introduction, the key distinction from the case of ordinary matter on a fixed de Sitter background is the presence of a gauge symmetry associated with the de Sitter isometry—namely, we are performing a dressing with respect to the observer. This parallels the situation in JT gravity, where the SL(2) isometry of the ambient AdS$_2$
  space becomes the gauge symmetry of the Schwarzian particle. Here, the observer plays the role of the Schwarzian particle, and our analysis closely follows that of the Schwarzian particle in \cite{Kitaev:2018wpr,Yang:2018gdb,Penington:2023dql}.\footnote{See also \cite{Jafferis:2019wkd,Kolchmeyer:2024fly}.}

\subsection{Brownian Motion}

First, let us temporarily disregard gauge symmetry and consider the Euclidean propagator of the observer, evolving from the initial configuration \(\lbrace x^{\mu}_0,0\rbrace\) to the final configuration \(\lbrace x^{\mu}_1,\mathbf{T}\rbrace\) on a Euclidean sphere \(S^D\). Here, \(x^{\mu}\) represents the observer’s position, while \(\mathbf{T}\) denotes the Euclidean clock time. The propagator is given by the path integral:
\begin{equation}
\int_{\text{Paths from } \lbrace x^{\mu}_0,0\rbrace \text{ to } \lbrace x^{\mu}_1,\mathbf{T}\rbrace }  
e^{- \int_0^1(m+\mathcal{E})\sqrt{\dot x^{\mu}(\lambda)\dot x_{\mu}(\lambda)}d\lambda-\int_0^1\mathbf{\dot{T}}\mathcal{E}d\lambda}
~{\mathcal{D}\mathcal{E}\mathcal{D}\mathbf{T}\mathcal{D}\mathbf{x}\over \text{Diff}(\mathbb{I})},
\end{equation}
where we have parametrized the particle’s trajectory as \(x^{\mu}(\lambda)\), mapping \([0,1]\) to \(S^D\), with boundary conditions \( x^{\mu}(0) =x^{\mu}_0\) and \(x^{\mu}(1) =x^{\mu}_1\). The clock path \(\mathbf{T}(\lambda)\) is constrained by the requirement that time flows forward, restricting the sum to monotonic maps from \([0,1]\) to \([0,\mathbf{T}]\). This constraint imposes the condition \(\mathbf{\dot{T}}(\lambda) > 0\).

The action possesses a reparametrization symmetry $\text{Diff}(\mathbb{I})$ of the form:
\begin{equation}
\lambda\rightarrow f(\lambda),~~~f(0)=0,~f(1)=1,~\dot f(\lambda)>0,
\end{equation}
which should be treated as a gauge symmetry. We can utilize this symmetry to fully fix the clock path:
\begin{equation}
\mathbf{T}(\lambda) = \mathbf{T}\lambda.
\end{equation}
This simply reflects the fact that we can always parametrize the particle’s trajectory using the clock time, which is one of the main reasons for introducing the clock. Upon this gauge fixing, the path integral over the clock energy variable \(\mathcal{E}\) reduces to a delta functional integral, enforcing the proper-time constraint:
\begin{equation}
\sqrt{\dot {\vec{x}}^2} = \mathbf{T}.
\end{equation}
This can be interpreted as fixing the induced metric of the particle along \(\tau\). Consequently, the observer’s path integral reduces to the following sum over all paths on the sphere:
\begin{equation}
\sum_{\text{Paths from } x^{\mu}_0 \text{ to } x^{\mu}_1} \delta(\sqrt{\dot {\vec{x}}^2}- \mathbf{T}) e^{-m \mathbf{T}}.
\end{equation}

At a microscopic level, summing over such paths (using an ultralocal measure, if we imagine discretizing the sphere) describes a Brownian motion on \(S^D\). The typical trajectory is a fractal with Hausdorff dimension 2. In the continuum limit, this path integral reduces to that of a standard nonrelativistic particle moving on \(S^D\), governed by the action\cite{Polyakov:1987ez}:
\begin{equation}\label{eqn:actobs}
-I_{\text{obs}} = - \int d \tau\frac{1}{2} \dot{\vec{x}}^2 +\mu .
\end{equation}
Here, \(\mu\) and \(\tau\) are the renormalized mass and time, and the Wiener measure for the nonrelativistic particle arises from summing over microscopic paths. Crucially, the clock time now serves as the time parameter for the nonrelativistic particle, rather than a naive measure of distance (or time) along \(S^D\). This is analogous to the hydrodynamic approach to black hole thermodynamics, where physical spacetime is mapped to a bulk ``fluid spacetime." In our case, the clock time plays the role of physical time, while \(S^D\) (or its analytic continuation \(dS_D\)) serves as the fluid spacetime.

In our application, we will be primarily interested in the regime where the observer is semiclassical, meaning its mass is large. In this limit, physical time and bulk time become proportional to each other. This can be seen by examining the classical action of a particle propagating over a short distance \(\Delta X\), which takes the form:
\begin{equation}
-\mu \Delta \tau - \frac{\Delta X^2}{2\Delta \tau}.
\end{equation}
Averaging over \(\Delta \tau\) slightly, the semiclassical propagator becomes peaked at \(\Delta X = \sqrt{2\mu} \Delta \tau\), reducing to the relativistic action:
\begin{equation}
-\sqrt{2\mu} \Delta X.
\end{equation}
This leads to the identification of the renormalized parameters \(\mu, \d \tau\) with the particle mass \(m\) and proper Eculidean time \(\d X\) in the continuum limit:
\begin{equation}
\mu = \frac{m^2}{2}, \quad \d\tau = \frac{\d X}{m}.
\end{equation}
That is, \(\d \tau\)  measures the proper time in units of the particle’s Compton wavelength, \(1/m\).
From now on, we will rescale $\d\tau\rightarrow{\d\tau\over m}$ to parametrize the particle's proper time.

Note that the phase space of the nonrelativistic particle is \(2D\)-dimensional, a property that remains valid under analytic continuation to \(dS_D\). The additional two dimensions, compared to a standard relativistic particle in \(dS_D\), arise from the inclusion of the clock.
In other words, before imposing the de Sitter constraint, the observer model is equivalent to a one-dimensional nonlinear sigma model with target space \(dS_D\) or \(S^D\).

Let's summarize the discussion so far in a simple case of $D$ dimensional flat space.

Before introducing the clock, the propagator of a relativistic particle in momentum space $\vec{p}=(p_0,p_1,...,p_d)$ takes the form  
\begin{equation}
G(\vec{x},0) = \int_0^{\infty} \d \tau \int \d \vec{p} \, e^{-(\vec{p}^2+m^2)\tau+\i \vec{x} \vec{p}} = \langle x | \frac{1}{\vec P^2 + m^2} | 0 \rangle.
\end{equation}
Here, \(\tau\) appears as a Schwinger proper-time parameter, which is integrated over. As a result, the propagator enforces the (Euclidean) on-shell condition \(\vec{P}^2 + m^2 = 0\), manifesting the standard relativistic dispersion relation.  

After introducing the clock, the structure of the path integral is modified. Rather than directly integrating over \(\tau\), we now consider  
\begin{equation}
G(\vec{x},\tau;\vec{0},0) = \int \d \vec{p} \, e^{- \frac{(\vec{p}^2+m^2)}{2 m}\tau+\i \vec{x}\vec{ p}} = \langle \vec{x} | e^{- \frac{(\vec{P}^2+m^2)}{2 m}\tau} | 0 \rangle.
\end{equation}
Here, \(\tau\) no longer acts as an auxiliary parameter to be integrated out, but instead represents the total clock time elapsed.   

To transition to Minkowski space, we can take analytic continuation ($x_0\rightarrow \i t,p_0\rightarrow\i E, i\in(1,...,d)$)\footnote{d=D-1.} of the propagator  
\begin{equation}
\int \d E \d p_i \, e^{-\i u \frac{(m^2 + p_i^2 - E^2)}{2m} + \i x_i p_i - \i E t}.
\end{equation}
Here \(u=-\i \tau\) is the Lorentzian proper time measured by the clock and therefore we can identify its Hamiltonian with the clock energy $\mathcal{E}$. On the other hand, $t$ is the Lorentzian bulk time so that $E$ is the energy measured by bulk observer. We see that these two energies are related by the following equation:  
\begin{equation}\label{eqn:H}
\mathcal{E} \equiv  \frac{E^2-m^2 - p_i^2}{2m},~~~E\sim m+{p_i^2\over 2m}+\mathcal{E},
\end{equation}
where the sign of the energy $\mathcal{E}$ is determined from \eqref{eqn:obsac}. Notice that a peculiar feature is that the system is evolved under $e^{\i \hat{\mathcal{E}} u}$ with respect to the clock time $u$ instead of the usual $e^{-\i \hat{\mathcal{E}}  u}$ evolution.
The equation \eqref{eqn:H} defines a new on-shell condition for the extended system, in which the energy of the clock is explicitly incorporated to contribute to the bulk energy $E$. We will use the generalization of this equation in section \ref{sec:obsHilbertspace}.

As a side note, recall that the on-shell condition \(\vec P^2 + m^2 = 0\) can be interpreted as the Hamiltonian constraint in one-dimensional quantum gravity along the particle's worldline. Here, we are arguing that introducing a clock modifies the constraint by including the clock energy, an effect that can be viewed as a simple illustration of \eqref{eqn:Hconint}.

\subsection{Two-sided Hilbert Space}\label{sec:obsHilbertspace}

We now turn to the imposition of constraints. Since we are dressing with respect to the observer, different embeddings of the observer’s trajectory in de Sitter space that are related by a de Sitter isometry should be regarded as equivalent. In other words, we must quotient out the \(SO(D,1)\) symmetry of the sigma model.

In order to construct the gauge invariant Hilbert space, it is necessary to introduce two particles, labeled \(L\) and \(R\). Classically, these correspond to two observers located at the two static patches of de Sitter space. The Hilbert space before imposing the constraint consists of wavefunctions \(\Psi(x_L^\mu, x_R^\mu)\) that depend on the two de Sitter coordinates.

\begin{figure}
    \centering
    \subfigure[]{\includegraphics[height=4.4cm]{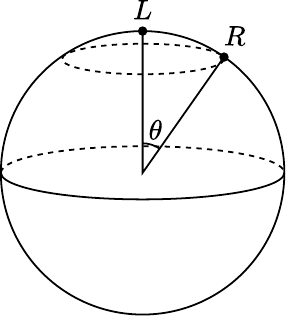}}\label{gauge_fix_euclidean}\qquad\qquad
    \subfigure[]{\includegraphics[height=4cm]{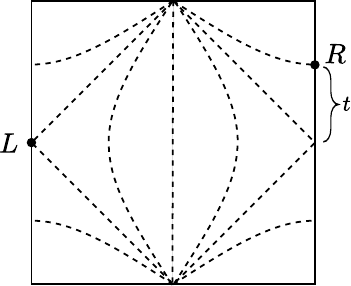}}\label{gauge_fix_lorentzian}
    \caption{The gauge fixing condition for the two observers in Euclidean sphere (a) and de Sitter space (b), the only degree of freedom left is the relative time shift ($t$) between the two observers.}
    \label{figure:gauge_fix}
\end{figure}

A natural way to impose the constraint is to restrict the wavefunction to depend only on gauge-invariant quantities, in this case, the geodesic distance between the two particles. In the Euclidean case $S^D$, this can be understood as follows: we first use the $SO(D+1)$ isometry to gauge-fix the left particle to the north pole of \(S^D\), and then use the residual \(SO(D)\) isometry to fix the position of the right particle, up to a relative angle between \(L\) and \(R\) (see Fig.~\ref{figure:gauge_fix} (a)). In the Lorentzian setting of dS$_D$, a similar procedure can be carried out: we first fix the left particle at the north pole of the spatial sphere \(S^{D-1}\) and at the \(t=0\) slice. However, the residual \(SO(D-1,1)\) isometry now generates multiple distinct orbits depending on the initial location of the the right particle, as illustrated in Fig.~\ref{figure:gauge_fix} (b).
In order to satisfy the charge constraint, we consider the situation where the two particles are antipodally located on the spatial sphere.\footnote{Otherwise one could always find a frame such that the two observers are in the same static patch, and they will not satisfy the corresponding boost constraint.} In this setup, the only remaining gauge-invariant quantity is the relative time shift \(t\) between the two particles. Consequently, the wavefunction reduces to an \(L^2\) function of \(t\).

More explicitly, in the embedding system \(Z = (Z_{-1}, Z_1, \dots, Z_D)\) for \(dS_D\), which satisfy the defining equation
\begin{equation}\label{eqn:emb}
-Z_{-1}^2 + Z_1^2 + Z_2^2 + \dots + Z_D^2 = 1.
\end{equation}
The locations of the two particles in these coordinates are given by
\begin{equation}\label{eqn:obsloc}
\begin{cases}
    Z^L = (0, -1, 0, 0, \dots, 0),\\
    Z^R = (\sinh t, \cosh t, 0, 0, \dots, 0).
\end{cases}
\end{equation}
From which we can also obtain the relation between $t$ and the geodesic distance $\ell$ of the two particles:
\begin{equation}
\cos\ell = Z^L \cdot Z^R = -\cosh t.
\end{equation}
This procedure reduces the phase space down to two dimensions, that can be interpreted as the relative time shift between the two clocks located at opposite static patches of de Sitter space and its canonical conjugate, as discussed in \cite{Chandrasekaran:2022cip}.

To complete the construction of the Hilbert space, we must determine its inner product. Later in section \ref{sec:coinv}, we will use the coinvariant formalism of \cite{Penington:2023dql} to derive the appropriate inner product. For now, we determine the measure by imposing the self-adjointness condition on the Hamiltonian.
Here, the Hamiltonian refers to the Hamiltonian of the nonlinear sigma model (or the clock Hamiltonian see \eqref{eqn:H}) restricting on the gauge invariant wavefunction. Since we have two particles, there are two Hamiltonians $\mathcal{E}_L$ and $\mathcal{E}_R$, later in section \ref{sec:coinv}, we will show that they acts the same on the gauge invariant Hilbert space\footnote{More precisely, the co-invariant Hilbert space.}. But for now, we simply focus on $\mathcal{E}_R$ that acts on wavefunctions that is only depend on $t$. In fact wavefunctions that depends on the geodesic distance are called spherical functions \cite{helgason2022groups}, and the Laplacian acting on such wavefunctions can be conveniently determined using the hyperbolic slicing, in which de Sitter space \( dS_D \) is foliated by constant \( Z_1 \) slices, corresponding to the orbits of \( SO(D-1,1) \):
\begin{equation}
Z_1 = \cosh t, \quad \{Z_{-1}, Z_2, \dots, Z_D\} = \sinh t \times \{ Y_{-1}, Y_2, \dots, Y_D \}.
\end{equation}
Here, \( Y \) denotes standard embedding coordinates for the spatial hyperbolic slice \( H_d \) with unit radius, so that the metric takes the form
\begin{equation}
\d s^2 = -\d t^2 + \sinh^2 t \, \d H_d^2.
\end{equation}
The Hamiltonian $\hat{\mathcal{E}}$, which is the Laplacian acting on wavefunctions that depend only on \( t \), is simply the radial component of the original Laplacian:
\begin{equation}\label{eqn:Hact}
\hat{\mathcal{E}} \psi(t) = -\frac{1}{2m} \left( \frac{1}{\sinh^d t} \frac{\d}{\d t} \sinh^d t \frac{\d}{\d t} + m^2 \right) \psi(t).
\end{equation}
Let's make a few comments about this formula. 
\begin{itemize}
    \item First, the Hamiltonian is symmetric under time reflection $\Theta$: \( \Theta \psi( t) = \psi^*(-t) \). In fact, this is a discrete gauge symmetry \cite{Susskind:2023rxm,Harlow:2023hjb} that has not yet been accounted for, and quotienting by it reduces the domain of \( t \) to the positive real line, \( \mathbb{R}_+ = [0, \infty) \).
    \item Second, we must impose a boundary condition at \( t=0 \). Since no additional particles are present at the south pole, we restrict attention to wavefunctions that are smooth at \( t=0 \).
    \item Finally, for the Hamiltonian to be self-adjoint, the inner product must take the form  
\begin{equation}\label{eqn:innobs}
\langle \psi | \phi \rangle = \int_0^{\infty} \d t \, \sinh^d t \, \psi^*(t) \phi(t).
\end{equation}
\end{itemize}

Having established the proper framework, we now solve for the spectrum of the Hamiltonian. The eigenvalues form a continuous spectrum $\mathcal{E}(s)$, parameterized by a real positive variable \( s \), and the eigensystem is given by
\begin{equation}\label{eqn:spectrum}
\mathcal{E}(s)= \frac{1}{2m} \left( \frac{d^2}{4} +s^2 -m^2  \right) ,\,\,
\langle t | s \rangle \equiv \psi_s(t) = {}_2F_1 \left( \frac{d}{2} + \i s, \frac{d}{2} - \i s, \frac{d+1}{2}; -\sinh^2 \frac{t}{2} \right).
\end{equation}
Here the eigenstate \( {}_2F_1(\alpha, \beta, \gamma; z) \) is the hypergeometric function, which satisfies \( {}_2F_1(\alpha, \beta, \gamma; 0) = 1 \). If we expand the wavefunction at large $t$, the wavefunction oscillates as $e^{\pm\i s t -d t}$, and we can think of $s$ as the bulk energy measured by time $t$. Then the energy relation $\mathcal{E}(s)$ can be thought of as the generalization of the on-shell equation \eqref{eqn:H} to de Sitter space. We will be interested in the semiclassical region where the mass of the particle is much larger than the clock energy $s\sim m \gg \mathcal{E}$, in this limit, we have the simplified energy relation:
\be\label{eqn:ESrelation}
\mathcal{E}(s)\sim s-m.
\ee
Since \( \hat{\mathcal{E}} \) is self-adjoint, the eigenfunctions satisfy the orthogonality condition and completeness relation:
\begin{equation}
\langle s | s' \rangle = \int_0^\infty \d t \, \sinh^d t \, \psi_s^*(t) \psi_{s'}(t) = \frac{\delta(s - s')}{\rho(s)},~~~\frac{\delta(t - t')}{\sinh^d t} = \int_0^\infty \d s \, \rho(s) \langle t | s \rangle \langle s | t' \rangle,
\end{equation}
where the Plancherel measure is given by (see appendix \ref{app:planch})
\begin{equation}
\rho(s) = \frac{2^{1-d} s \Gamma \left( \frac{d}{2} - \i s \right) \Gamma \left( \frac{d}{2} + \i s \right) \sinh (\pi s)}{\pi \Gamma^2 \left( \frac{d+1}{2} \right)}.
\end{equation}
For special cases of \( dS_2 \) and \( dS_3 \), and for large \( s \), the measure simplifies to
\begin{equation}
\rho(s) =
\begin{cases}
    s \tanh (\pi s), \quad d=1, \\
    \frac{2s^2}{\pi}, \quad d=2.
\end{cases}~~~~
\rho(s \to \infty) \approx \frac{2^{1-d}}{\Gamma \left( \frac{1+d}{2} \right)^2} s^d.
\end{equation}

We now apply these results to analyze the sphere partition function for the observer studied in \cite{Maldacena:2024spf}. Up to an overall phase, this partition function can be interpreted as the inner product of the Hartle-Hawking (HH) wavefunction with itself. 
The Hartle-Hawking wavefunction admits an expansion in the energy basis $|s\rangle$:
\begin{equation}
|\text{HH}\rangle = \int \mathrm{d}s \, \rho_{\text{HH}}(s) |s\rangle, \quad
\langle \text{HH}|\text{HH}\rangle = \int \mathrm{d}s \, \frac{\rho_{\text{HH}}(s)^2}{\rho(s)}.
\end{equation}
The density $\rho_{\text{HH}}(s)$ is determined by matching with the large-$m$ behavior of the sphere partition function \cite{Maldacena:2024spf}:
\begin{equation}
\i Z_{\text{sphere}}' \sim \frac{2\pi e^{-2\pi m} m^d}{d!} \int_0^\infty \mathrm{d}\mathcal{E} \, e^{-2\pi \mathcal{E} + S_{\text{clock}}+S_{\text{dS}}},
\end{equation}
where the clock energy $\mathcal{E}$ should be identified with $\mathcal{E}(s)$. The $e^{S_{\text{clock}}}=\int {\d \mathcal{T}_0\over 2\pi}$ comes from the zero mode integral of the clock time, since here we are not integrating over such zero mode, we will set it to be ${1\over 2\pi}$. The $S_{\text{dS}}$ is the de Sitter entropy that we include as an overall normalization factor. This matching fixes the form of the Hartle-Hawking wavefunction:
\begin{equation}\label{eqn:HHwave}
|\text{HH}\rangle
\sim \frac{2^{\frac{1-d}{2}} m^d}{\Gamma\left(\frac{1+d}{2}\right)\sqrt{d!}} e^{\frac{S_{\text{dS}}}{2}} \int_m^\infty \mathrm{d}s \, e^{-\pi s} |s\rangle.
\end{equation}
Notice that the $e^{-\pi s}$ energy dependence is consistent with expression (2.13) in \cite{Chandrasekaran:2022cip}.
Having the wavefunction, we can investigate the time evolution of the Hartle-Hawking state. A simple example is the expectation value of the time variable $t$, which under evolution by $\mathcal{T}$ is given by :
\begin{equation}
\langle \text{HH}| e^{-i \hat{\mathcal{E}} \mathcal{T}} \, \hat{t} \, e^{i \hat{\mathcal{E}} \mathcal{T}} |\text{HH}\rangle 
\sim \frac{ m^d}{d!} e^{S_{\text{dS}}} \int \mathrm{d}s \, \mathrm{d}s' \, e^{-\pi s - \pi s' + i(s-s')t - i \mathcal{T}(s-s')} \, t 
\sim \mathcal{T} \langle \text{HH}|\text{HH}\rangle.
\end{equation}
The linear growth in $\mathcal{T}$  of the relative time separation indicates that the clock is functioning properly, faithfully tracking the observer's proper time evolution. This behavior also implies that the time-evolved Hartle-Hawking states are nearly orthogonal. Recall that the Lorentzian partition function \eqref{eqn:ZTdef} is defined by inserting finite clock time evolution in the sphere partition function, this corresponds to the overlap between the time-evolved Hartle-Hawking state and the original Hartle-Hawking state, which is the return amplitude of the Hartle-Hawking wavefunction under clock-time $\mathcal{T}$ evolution. Using \eqref{eqn:HHwave}, this is given by:
\begin{equation}\label{eqn:Zltime}
\i Z(i\mathcal{T}) = \langle \text{HH}| e^{i\hat{\mathcal{E}}\mathcal{T}} |\text{HH}\rangle \sim \frac{m^d}{d!} e^{S_{\text{dS}}} \int \mathrm{d}\mathcal{E}\, e^{-2\pi(m + \mathcal{E}) + i\mathcal{E}\mathcal{T}}.
\end{equation}
Alternatively, with the insertion of a window function $f$ \eqref{eqn:smear}, we have:
\begin{equation}\label{eqn:Zltime_f}
\i Z_f(i\mathcal{T}) \sim \frac{ m^d}{d!} e^{S_{\text{dS}}} \int \mathrm{d}\mathcal{E}\, e^{-2\pi(m + \mathcal{E}) + i\mathcal{E}\mathcal{T}} f(\mathcal{E}) \sim \frac{ m^d}{d!} e^{S_{\text{dS}} - 2\pi(m + \mathcal{E}_*) + i\mathcal{T}\mathcal{E}_* - \frac{\sigma^2\mathcal{T}^2}{2}} \approx 0,
\end{equation}
which exhibits exponential decay, vanishing after a timescale of order $\sqrt{S_{\text{dS}}}$. 

In the above calculation, we view the Lorentzian partition function as the expectation value of the clock evolution operator in the Hartle-Hawking state, whose form was determined via saddle point approximation of all other degrees of freedom, and so the geometry is just a sphere. An alternative way to derive the result \eqref{eqn:Zltime_f} is to use saddle point approximation as in \cite{Maldacena:2024spf}, all the integral will be the same expect the clock partition function, due to the different boundary conditions. This net Lorentzian evolution of the clock modifies the clock partition function into:
\be
Z_{\text{clock}}=\int {\d \mathcal{E}_0\d\mathcal{T}_0\over 2\pi} e^{\i \mathcal{T}\mathcal{E}_0-2\pi \mathcal{E}_0},
\ee
that after insertion into the sphere partition function leads to \eqref{eqn:Zltime}. One might wonder whether the sphere partition function of the observer could be more directly related to a conventional trace of the nonlinear sigma model, this however is not straightforward. Even in the Schwarzian case, the disk partition function could not be directly expressed as a conventional trace of the Schwarzian particle, and requires the use of a certain regularized trace defined in \cite{Kitaev:2018wpr}.

We could also use this framework to calculate higher-point dressed correlation functions, as a generalization of the Schwarzian calculation \cite{Kitaev:2018wpr,Yang:2018gdb,Kolchmeyer:2024fly}. But for the current paper, we will focus on the double cone correction in de Sitter space that contributes nondecaying behavior of the SFF. In such case, we will need to use
this observer's Hilbert space to compute the observer's partition function on the double cone geometry, see Section~\ref{sec:obsdc}.

\subsection{The Coinvariant Approach}\label{sec:coinv}

We now describe a more systematic approach of imposing the gauge constraint on the Hilbert space. Instead of fixing the gauge explicitly, we adopt the coinvariant formalism discussed in \cite{Penington:2023dql}, which provides a rigorous framework for defining the physical Hilbert space. The essential idea is to introduce an equivalence relation in the Hilbert space, identifying any two states that are related by a gauge transformation as physically equivalent. The inner product on the space of coinvariants is then defined via integration over the gauge group \( G \) using the Haar measure:
\begin{equation}\label{eqn:coinv}
\langle \Psi'|\Psi\rangle = \int_G \d g \, ( \Psi', R(g) \Psi),
\end{equation}
where \( (, ) \) denotes the inner product in the original Hilbert space, and \( \langle , \rangle \) represents the inner product in the coinvariant Hilbert space.

For the case of interest, we can always fix the locations of the two particles as in \eqref{eqn:obsloc}. This implies that the space of coinvariants is spanned by wavefunctions of the form
\begin{equation}\label{eqn:coinwav}
\Psi(Z^L, Z^R) = \delta(Z^L_{i\neq 1}) \delta( Y^R_{i\neq -1}) \psi(t),
\end{equation}
where the first delta function enforces the gauge fixing of the left particle, while the second fixes the location of the right particle. The inner product \eqref{eqn:coinv} then takes the explicit form:
\begin{equation}\label{eqn:coinvinn}
\langle \psi | \phi \rangle = \int_G \d g \int \d Z^L \d Z^R \psi^{*}(t) \phi(t) \delta(Z^L_{i\neq 1}) \delta(Y^R_{i\neq -1}) \delta(gZ^L_{i\neq 1}) \delta(gY^R_{i\neq -1}) \delta(Z^L \cdot Z^L -1) \delta(Z^R \cdot Z^R -1),
\end{equation}
where \( G \) is the gauge group \( SO(D,1) \). It is straightforward to show that this expression reduces to the previously obtained inner product \eqref{eqn:innobs}.

To complete the construction, we now determine the Hamiltonian in this framework. The procedure is to first compute the Hamiltonian and constraint operators in the original two-particle Hilbert space and then restrict to the physical subspace. For simplicity, we illustrate the approach in \( dS_2 \), and the generalization to higher dimensions is straightforward. %with generalization to higher dimensions discussed in Appendix \ref{app:Hcon}.

The embedding coordinates for \( dS_2 \) can be parametrized as:
\begin{equation}
Z_{-1} = \sinh t \cosh \rho, \quad Z_{1} = \cosh t, \quad Z_2 = \sinh t \sinh \rho.
\end{equation}
where SL(2) isometry acts as:
\begin{equation}
\Lambda_0 = \sinh \rho \, \partial_t - \coth t \cosh \rho \, \partial_{\rho}, \quad
\Lambda_1 = \cosh \rho \, \partial_t - \coth t \sinh \rho \, \partial_{\rho}, \quad
\Lambda_2 = \partial_{\rho},
\end{equation}
that satisfying the algebra:
\begin{equation}
[\Lambda_0, \Lambda_1] = \Lambda_2, \quad [\Lambda_0, \Lambda_2] = -\Lambda_1, \quad [\Lambda_1, \Lambda_2] = -\Lambda_0.
\end{equation}
The two observers occupy different patches of global de Sitter space. Their coordinate relations are:
\begin{equation}
t_L = \i \pi - t, \quad \rho_L = \rho; \quad t_R = t, \quad \rho_R = \rho.
\end{equation}
In this notation, the wavefunction \eqref{eqn:coinwav} takes the form:
\begin{equation}\label{eqn:coinwavds2}
\Psi = \frac{1}{t_L} \delta(t_L) \delta(\rho_L) \delta(\rho_R) \psi(t_R).
\end{equation}
In Lorentzian signature, the observer's action \eqref{eqn:actobs} takes the form:
\begin{equation}
-I_{\text{obs}} = \i \int \d u \left[ \frac{m}{2} \left(-\dot{t}_L^2 + \sinh^2 t_L \dot{\rho}_L^2 - 1 \right) + \frac{m}{2} \left(-\dot{t}_R^2 + \sinh^2 t_R \dot{\rho}_R^2 - 1 \right) \right].
\end{equation}
From this, we can determine the canonical momenta and Hamiltonians for the two particles:
\begin{align}
p_{t_{L,R}} &= -m\dot{t}_{L,R}=-\i \partial_{t_{L,R}}, \quad p_{\rho_{L,R}} = m\sinh^2 t_{L,R} \dot{\rho}_{L,R}-\i \partial_{\rho_{L,R}}, \\
\mathcal{E}_{L,R} &=\frac{1}{2m \sinh t_{L,R}} p_{t_{L,R}} \sinh t_{L,R} p_{t_{L,R}} - \frac{1}{2m\sinh^2 t_{L,R}} p_{\rho_{L,R}}^2 - {m\over 2}.
\end{align}
Here, we have chosen an operator ordering such that \( \mathcal{E}_{L,R} \) coincides with the standard Laplacian $ {\Lambda_0^2-\Lambda_1^2-\Lambda_2^2-m^2\over 2 m}$ on \( dS_2 \).
The SL(2) constraints acting on both particles take the form:
\begin{align}\label{eqn:Hcon}
H &= \sinh \rho_R p_{t_R} - \coth t_R \cosh \rho_R p_{\rho_R} - \sinh \rho_L p_{t_L} + \coth t_L \cosh \rho_L p_{\rho_L}, \\
K  &= \cosh \rho_R p_{t_R} - \coth t_R \sinh \rho_R p_{\rho_R} - \cosh \rho_L p_{t_L} + \coth t_L \sinh \rho_L p_{\rho_L}, \\
P &= p_{\rho_R} + p_{\rho_L}.\label{eqn:Pcon}
\end{align}
These constraints must annihilate the wavefunction in the coinvariant Hilbert space, that gives the equivalence relation:
\begin{equation}\label{eqn:coin_equiv_relation}
    -p_{\rho_L} = p_{\rho_R}= {\tanh t_L \tanh t_R \sinh(\rho_R-\rho_L)\over \tanh t_R+\tanh t_L \cosh(\rho_L-\rho_R)} p_{t_R},\quad p_{t_L} = {\tanh t_L+\tanh t_R \cosh(\rho_L-\rho_R)\over \tanh t_R+\tanh t_L \cosh(\rho_L-\rho_R)}p_{t_R}.
\end{equation}
Substituting these relations into the Hamiltonian when acting on wavefunction of the form \eqref{eqn:coinwavds2}, we find that both \( \mathcal{E}_L \) and \(\mathcal{E}_R \) act in the same way as $\hat{\mathcal{E}}$ on \( \Psi \) (for \( d=1 \) in \eqref{eqn:Hact}), as claimed.\footnote{The easy way to see \( \mathcal{E}_L \)=\( \mathcal{E}_R \) is to notice that their difference is a linear combination of the constraints \eqref{eqn:Hcon}-\eqref{eqn:Pcon}.}

In the end, we comment that we could also include matters in this framework as in \cite{Penington:2023dql}, where we use the same gauge fixing condition but then $\psi(t)$ will be evaluated in $\mathcal{H}_{\text{QFT}}$. The constraint of the residual $SO(d)$ symmetry will need to be imposed on $\mathcal{H}_{\text{QFT}}$, but since its a compact group, we will simply require the matter states to be rotation invariant. In other words, the whole bulk Hilbert space is given by 
\be\label{eqn:Hbulk}
\mathcal{H}_{\text{bulk}}=L^2(\mathbb{R}_+)\otimes \tilde{\mathcal{H}}_{\text{QFT}}, \quad \tilde{\mathcal{H}}_{\text{QFT}}={\mathcal{H}_{\text{QFT}}\over SO(d)}.
\ee
The Hamiltonian in the presence of matter excitations will now act differently due to the inclusion of matter charges, but we will not need that in this paper as for the late time double cone geometry, the matter state will be in a ground state due to quasi-normal mode decay and the Hilbert space reduces to $L^2(\mathbb{R}_+).$

\section{Double cone geometry}\label{sec:doublecone}

In this section, we explore the wormhole correction to the Lorentzian partition function. As demonstrated in the previous section, when considering a single copy of the system, the Lorentzian partition function exhibits exponential decay. However, for a system with finite entropy, such decay cannot persist indefinitely. This observation was highlighted by \cite{Dyson:2002pf,Goheer:2002vf}, who argued that the decay of the matter two-point function—or more generally, the full de Sitter isometry group—is inconsistent with the finiteness of de Sitter entropy.

A similar paradox emerges in the context of black hole systems, as first pointed out by \cite{Maldacena:2001kr}. \cite{Maldacena:2001kr} argued that, due to the discreteness of the black hole energy spectrum, the two-point function should eventually saturate to a small, non-perturbative value of order \( e^{-S_{\text{BH}}} \), where \( S_{\text{BH}} \) is the black hole entropy. However, perturbative calculations alone fail to capture this saturation, resulting in a paradox. Recently, this paradox was addressed by Saad-Shenker-Stanford, who showed that wormhole effects can induce non-decaying behavior in the two-point function \cite{Saad:2018bqo,Saad:2019pqd}. 

This discovery is deeply connected to recent progress in understanding the random matrix behavior of many-body chaotic systems, such as black holes. In these developments, the spectral form factor serves as a key diagnostic to probe the underlying random matrix spectral statistics of such systems. 

Given these developments, it is natural to ask whether the wormhole solution proposed by \cite{Saad:2018bqo} can also resolve the analogous puzzle in de Sitter space. In this section, we will argue that this is indeed the case, at least for the leading geometry known as the double cone geometry. The double cone is a dominant wormhole configuration that contributes to the ``ramp" of the spectral form factor in black hole systems. It is a cylindrical wormhole geometry that connects two asymptotic boundaries. 

In the de Sitter context, constructing such a wormhole geometry requires considering configurations involving two observers whose clocks run with opposite total time periodicity. From a putative microscopic perspective, this setup corresponds to calculating the return probability—defined as the absolute value squared of the return amplitude—for a system evolved over a clock time \( \mathcal{T} \), that is referred to as SFF in \eqref{eqn:dSSFF}.

\subsection{Double cone saddle}
\begin{figure}[h]
    \centering
    \includegraphics[width=0.25\textwidth]{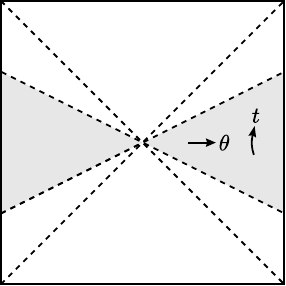}
    \caption{The double cone geometry is obtained by taking a finite quotient of the static de Sitter spacetime.}
    \label{im:double_cone_penrose_diagram}
\end{figure}

The double cone geometry (see Fig.~\ref{im:double_cone_penrose_diagram}), introduced in \cite{Saad:2018bqo}, is a universal complexified on-shell solution associated with geometries with killing horizon, obtained by taking a finite quotient with respect to the boost symmetry.

The reason the geometry is complexified is that the quotient produces a conical singularity at the horizon, and in order to avoid the singularity, the radial coordinate \(\rho\) is analytically continued from \(0^-\) to \(0^+\) through the lower half-plane. Equivalently, the local metric is given by
\be\label{eqn:dcgluing}
\d s^2 = - (\rho - \i\epsilon)^2\, \d t + \d \rho^2 + \d s_{\Sigma}^2,\quad t\sim t+T,
\ee
where $T$ is the parameter of the double cone that is given by the amount of boost evolution of the quotient.
The choice of the minus sign ensures that the time periodicity acquires a small positive Euclidean component, so that \(\i T \to \i T+ \epsilon\).
For our case, we consider the double cone geometry constructed from vaccum de Sitter space dS$_D$ by taking a quotient of the boost symmetry with repsect to the static patch,  the metric is given by:
\be
\d s^2=-\sin^2(\theta-\i\epsilon) \d t^2+\d\theta^2+\cos^2\theta \d s^2_{\text{S}_{d-1}},~~~t\sim t+T,~~~\theta\in[-{\pi\over 2},{\pi\over 2}].
\ee
Such a geometry contains two closed timelike geodesics at the pode ($\theta={\pi\over 2}$) and antipode ($\theta=-{\pi\over 2}$), where a pair of observers can live, and therefore contributes as a connected geometry in the summation of the SFF in \eqref{eqn:dSSFF}. Using  saddle point approximation, this reduces the path integral \eqref{eqn:dSSFF} into a summation over double cone geometries with different periodicities:
\be
\text{SFF}\approx \sum_{\text{Double cones}} e^{-I_{\text{EH}} - I_{\text{obs,L}} - I_{\text{obs,R}}} \Big|_{\Delta \mathcal{T}_R = \mathcal{T}, \Delta \mathcal{T}_L = -\mathcal{T}}.
\ee
It is easy to see that the on-shell action $I_{\text{EH}}$ of the double cone vanishes due to the analytic continuation procedure:
\be
-I_{\text{EH}}={(D-1)\over 8\pi G_N}\int \sqrt{g} =0.
\ee
Therefore, the path integral reduces to summation of one-loop determinants:
\be\label{eqn:DCSum}
\text{SFF}\approx \sum_{\text{Double cones}} Z_{\text{DC,matter}}Z_{\text{DC,obs}}
\ee
Here $Z_{\text{DC,matter}}$ represents the matter partition functions (including gravitons) on the double cone geometry, that will be discussed in Section \ref{sec:dcm}, and $Z_{\text{DC,obs}}$ represents that observer partition function, which will be the subject of Section \ref{sec:obsdc}. Our goal is then to perform an explicit evaluation of both terms.

Before we move on, we give several comments about the double cone wormhole, some of them are special to de Sitter space: 
\begin{itemize}
        \item Due to the analytic continuation of the radial coordinate, the metric remains smooth across the entire manifold. Importantly, due to its analyticity, physical quantities derived from the geometry are independent of the specific value of \(\epsilon\). This is because the contour of \(\rho\) can always be deformed, ensuring that the physical results are invariant under variations in \(\epsilon\).
    \item The vanishing of on-shell Einstein-Hilbert action means that the double cone is suppressed by a factor of \(e^{-2S_{\text{dS}}}\) relative to the disconnected sphere geometry. This comes from the absence of contractable time circles in the double cone.
    \item The double cone construction is universal for any Killing horizon due to the universal near-horizon Rindler structure $\mathbb{R}_2\times \Sigma$. This is the reason why it applies to both the black hole and de Sitter horizons.
    \item In de Sitter spacetime, there exists a larger family of double cone geometries constructed from Schwarzschild-de Sitter geometry. In this case, both the black hole and cosmic horizons require gluing as in \eqref{eqn:dcgluing} and the wormhole will take the topology of $T^2\times S^2$. We think that such geometries should be included in the summation of \eqref{eqn:DCSum}. However, as we will argue in Section~\ref{sec:obsdc}, the observers' partition function on these geometries exhibits quasi-normal mode decay and therefore does not contribute at late times. 
    An alternative perspective is to consider the quasi-normal mode decay as originating from black holes (when the black holes are small such that the observer will not fall into the black holes): these can be viewed as matter excitations on vacuum de Sitter space after gauge-fixing the observer at the center of the static patch. Consequently, they should follow the same quasi-normal mode decay behavior as ordinary matter fields.
    \item The analytic continuation (and consequently the complex metric) provides a way to evade several no-go theorems in Riemannian geometry that would otherwise exclude those wormhole geometries. A simple example appears in two dimensions: a naive toroidal wormhole with a positive cosmological constant would be ruled out by the Gauss-Bonnet theorem, which requires  
\be
\int \sqrt{g} R = 0,
\ee
in apparent contradiction with \( R > 0 \). However, the double cone geometry circumvents this constraint by allowing for negative volume.  
Explicitly, consider the two-dimensional de Sitter space, \( dS_2 \), where the double cone metric takes the form  
\be\label{eqn:ds2dc}
ds^2 = -\sin^2\theta \, dt^2 + d\theta^2, \quad \theta \sim \theta+2\pi, \quad t\sim t+T.
\ee 
This geometry has the topology of a torus, seemingly contradicting the Gauss-Bonnet theorem. However, an explicit evaluation of the Einstein-Hilbert action gives  
\be
\int \sqrt{g} R = 2 \int dt \int d\theta \, i \sin\theta = 0.
\ee
Here, the integral is computed via analytic continuation on the complexified manifold. This confirms that the \( dS_2 \) double cone satisfies the Gauss-Bonnet constraint despite its positive curvature. The crucial point is that in a complexified manifold, the volume needs not be positive definite.

    \item Another example is the Synge theorem that an orientable Riemanian manifold with positive sectional curvature must be simply connected in even dimensions \cite{petersen2006riemannian}.The idea behind this theorem is that on such a manifold, the nonrelativistic particle action along any closed geodesic necessarily has negative modes. If the fundamental group were nontrivial, there would exist a noncontractible closed geodesic that locally minimizes the action, leading to a contradiction. A torus, for example, has a nontrivial fundamental group and clearly does not satisfy the theorem. 
    
    In the case of a double cone geometry, the usual obstruction is avoided due to the presence of singular geodesics.  For instance, in the \(dS_2\) double cone geometry \eqref{eqn:ds2dc}, the minimum geodesics are located at the singularity \(\theta = 0,\pi\), which are circumvented by the $\i \epsilon$ prescription. Nevertheless, this issue persists in the context of Euclidean wormholes, obstructing constructions such as the generalization of the double trumpet in de Sitter space.

    \item The de Sitter double cone geometry without the presence of observers was discussed in the context of cosmological wavefunctions, where it was proposed to provide non-perturbative corrections to the universe's density matrix as a bra-ket wormhole \cite{Chen:2020tes,Fumagalli:2024msi}. It will be interesting to understand its connection with the current story.
\end{itemize}

\subsection{Matter partition function}\label{sec:dcm}
In this section, we examine the ordinary matter partition function on the double cone geometry. This analysis closely parallels the discussion in \cite{Chen:2023hra}, as their general arguments are equally applicable to the de Sitter case. We include this section primarily for self-consistency and to ensure clarity.

The main idea is that the one-loop matter partition function on the double cone can be interpreted as evaluating a bulk trace:
\be\label{eqn:QFTtr}
Z_{\text{DC,matter}}(T)=\Tr_{\text{bulk}} e^{-\i \tilde K_m T},~~~\tilde K_m=K_m-\i \epsilon H_m.
\ee
Here both $K_m$ and $H_m$ are operators acting on the matter Hilbert space that defines the bulk trace. 
$K_m$ is the boost operator associated with the de Sitter horizon, while 
$H_m$ is the Minkowski Hamiltonian defined locally at the horizon. To understand the modified boost operator $\tilde K_m$, note that 
$K_m$ is non-vanishing everywhere except at the bifurcation center. Thus, 
$\tilde K_m$ generally acts as $K_m$. However, at the bifurcation center, where 
$K_m$ vanishes, $\tilde K_m$ acts as a small Minkowski time translation in the imaginary direction. This introduces the $-\i \epsilon$ regularization in \eqref{eqn:dcgluing}. Since the geometry at the bifurcation center is locally Rindler, such a local symmetry operator 
$H_m$ can always be constructed. As shown in \cite{Chen:2023hra}, the eigenvalues of 
$\tilde K_m$ are given by the quasi-normal frequencies of the thermofield double geometry.
As an example, consider a scalar field 
$\phi$, and examine its Klein-Gordon equation near the horizon using the near-horizon metric \eqref{eqn:dcgluing}. The transverse direction decouples, and the wave equation effectively reduces down to two dimensions under the separation of variables:
\begin{equation}\label{eqn:nearhorizon equation}
\rho \partial_\rho \big( \rho \partial_\rho \phi \big) -\partial_{t}^2 \phi = 0,
\end{equation}
where $\phi$ depends only on $(t,\rho)$.
The equation \eqref{eqn:nearhorizon equation} admits separable solutions of the form:
\be\label{eqn:dcbdy}
\phi(\rho)= a(k) e^{-\i k t+\i k\log \rho}+b(k) e^{-\i k t-\i k \log \rho},
\ee
where $k$ remains to be determined by the gluing condition across the double cone. When we analytically continue $\rho$ through the lower half-plane  $\rho\rightarrow e^{-\i \pi} \rho$ as prescribed by the double cone regularization, the ratio ${a_k\over b_k}$ acquires an overall factor $e^{2\pi k}$. However, this ratio is uniquely determined by the smoothness condition at the pode and antipode. In order to be consist with the gluing condition, this requires either  $a_k$  or $b_k$ to vanish, or if $k=\i n $ the absence of $\log \rho$ term. Such condition determines the frequency $k$ has to be either quasi-normal frequency or anti-quasi-normal frequency \cite{Chen:2023hra}. The possibility of anti-quasi-normal frequency is ruled out by the requirement that  $H_m$ be positive definite, ensuring that the eigenvalues of  $\tilde K_m$  have a negative imaginary part. Taken together, this means that the bulk trace takes the form of a production of quasi-normal frequencies:
\be\label{eqn:ZmQNM}
Z_{\text{DC,matter}}(T)=\prod_{\omega_{\text{QNM}}}{1+e^{-\i \omega_f T}\over 1-e^{-\i \omega_b T}},
\ee
where $\omega_f$ and $\omega_b$ are the quasi-normal frequencies for fermions and bosons, respectively, and we have used the notation that when there are degeneracies $N_{\omega}$, the product is taken $N_{\omega}$ times. 
To understand this formula, let's first notice that the matter partition function contains poles and zeros at $T={2\pi n\over \omega_b}$ and $T={2\pi (n+1/2)\over \omega_f}$ for $n\in \mathbb{Z}$. Such structures come from the appearance of zero modes when we analytic continue $T$ such that the frequencies $k=\pm{2\pi n\over T}, n\neq 0$ (take bosonic field as an example) matches with the quasi-normal frequencies $\omega$ and anti-quasi-normal frequencies $\bar\omega=-\omega$. Just like in the case of harmonic oscillator, for each quai-normal frequency $\omega$, that takes the infinite product form of:
\be\label{eqn:infprod}
\prod_{n> 0} {1\over ({2\pi n\over T})^2-\omega^2}={\omega\over 2\pi}{1\over \sin({T\omega\over 2})}\times e^{\# T}.
\ee 
To determine the overall normalization including the ground state energy, we can take $T\rightarrow \infty$, in which limit the modified boost evolution becomes a projection to the Hartle-Hawking state. Since the Hartle-Hawking state is invariant under the boost symmetry, $Z_{\text{DC,matter}}(T)$ should approach to one. That leads to \eqref{eqn:ZmQNM}, under the assumption that $Z_{\text{DC,matter}}(T)$ is a meromorphic function of $T$.
We note that the assumption that $Z_{\text{DC,matter}}(T)$ being a meromorphic function assumes that the subtly issue of phases coming from the contour rotation in the path integral of negative modes that was important in the discussion of the sphere partition function can be accounted via analytic continuation of $T$. On the double cone geometry this seems to be a reasonable assumption due to the bulk interpretation that the partition function evaluates a trace in the bulk matter Hilbert space \eqref{eqn:QFTtr} and the property that the system is stable. In addition, one can also directly evaluate the partition function for negative mass squared fields for specific examples such as dS$_2$ double cone (see Appendix~\ref{app:scalardS2}) and find that the final answer is indeed positive definite.

In the above formula, we didn't impose the $SO(d)$ gauge constraint as in \eqref{eqn:Hbulk}, that could be imposed by hand by restricting to the rotation invariant quasi-normal modes in the product of \eqref{eqn:ZmQNM}. This does not change the late time behavior as the leading quasi-normal frequency is given by a rotation invariant mode.
At late times, this one-loop partition function decays to one, with the decay rate governed by the quasi-normal frequencies with the smallest imaginary part. For example, in the case of scalar field in dS$_{d+1}$ the quasi-normal frequency is given by:
\be\label{eqn:QNM}
\omega_{n,\pm}=\pm\nu-\i (n+{d\over 2}),\quad\bar\omega_{n,\mp}=\pm \nu+\i (n+{d\over 2});\quad n\geq 0, \quad \nu = \sqrt{m^2 - {d^2\over 4}}.
\ee
If we consider the bulk theory contains only massive particles, the late time behavior of the one-loop partition function takes the form:
\be
Z_{\text{DC,matter}}(T)\sim 1+ 2e^{-{d\over 2}T} \cos(\nu T).
\ee
So that it approaches to one after a few Hubble times. Notice that on the double cone, there are two closed geodesic circles (pode and antipode) that the matter particle can wind around, this explains the two branches of the cosine factor. More generally, in the presence of massless particles, the lowest quasi-normal frequency of the system is dimensional dependent. For instance for $dS_4$, the lowest quasi-normal frequency is given by the spin one particle, that decays as $e^{-T}$ \cite{Anninos:2020hfj}, indicating the Thouless time of the system, defined by the time when $Z_{\text{DC,matter}}(T)$ goes to a constant, is given by the Hubble time.

\subsection{Observers in the double cone}\label{sec:obsdc}

Next, we will discuss the observer's path integral in the double cone geometry, which differs from ordinary matter path integral discussed above due to gauge constraint as discussed in Section \ref{sec:Obs}. 
In other words, the double cone is defined relative to the observer, and we’ll see that this dressing is crucial for the double cone to contribute at late times. Without it, the double cone’s contribution decays exponentially in time, as a consequence of the quasi-normal modes of the observer.

First, we will analyze the double cone's contributions using the observer's coinvariant Hilbert space (from the canonical quantization framework in Section \ref{sec:Obs}). This is in parallel with the discussion of the standard trace of wormhole Hilbert space in Section 5.1 of \cite{Chen:2023hra}. The actual trace on the double cone geometry differs from this standard trace by compactification of the relative time shift, as we will discuss later. The purpose of the current discussion is to address the issue of quasi-normal mode decay of the observer on the double cone.

In the black hole case, the wormhole Hilbert space in the limit of $G_N \rightarrow 0$ factorizes as $\mathcal{H}_{\text{QFT}} \otimes L^2(\mathbb{R})$, where the $L^2(\mathbb{R})$ part describes the gravitational Hilbert space corresponding to the relative time shift $T_{\text{rel}}$ and its canonical conjugate $m$ of the thermofield double. What is important in the current discussion is the trace of the $L^2(\mathbb{R})$ part, which takes the form (see equation (61) in \cite{Chen:2023hra}):
\begin{equation}\label{eqn:Zwh}
Z_{\text{wh}} \equiv \Tr_{L^2(\mathbb{R})} f(m)^2 = \frac{1}{2\pi} \int_{-\infty}^{\infty} \! \mathrm{d} T_{\text{rel}} \int \! \mathrm{d} m \, f(m)^2,
\end{equation}
with insertion of some window function $f(m)$.

In the de Sitter case, the analog of the gravitational Hilbert space is the observer's Hilbert space discussed in Section~\ref{sec:Obs}, which is $L^2(\mathbb{R}_+)$, where the restriction to $\mathbb{R}_+$ comes from the quotient of time reversal symmetry. In fact, it is already shown in \cite{Chandrasekaran:2022cip} that the de Sitter Hilbert space with observers present in both static patches takes the form of $\mathcal{H}_{\text{QFT}} \otimes L^2(\mathbb{R}_+)$. Our analysis differs slightly from that of \cite{Chandrasekaran:2022cip} in the sense that \cite{Chandrasekaran:2022cip} directly fixes the clock in the center of the static patch, and consequently only considers imposing the boost constraint. In our analysis, we do not fix the position of the clock and must take into account the full de Sitter isometry as well as the time reversal symmetry. In particular, the restriction to $\mathbb{R}_+$ in \cite{Chandrasekaran:2022cip} comes from the constraint of the energy being positive rather than quotient the time reversal symmetry as in Section~\ref{sec:obsHilbertspace} , although such subtle difference do not show up after the compactification of the relative time.  In any case, if we view the double cone as evaluating trace over the co-invariant Hilbert space with insertion of the evolution of the observers boost operator $e^{-\i K_{\text{obs}} T}$, we get:
\be\label{eqn:dctr}
\Tr e^{-\i K_{\text{obs}} T} f(\mathcal{E})^2=\Tr f(\mathcal{E})^2=\int \d s \rho(s)\d t \sinh^{d} t\psi_s(t)^*\psi_{s}(t) f(s-m)^2\sim \int {\d s \d t\over 2\pi} f(s-m)^2.
\ee
We see that we get the same integral as \eqref{eqn:Zwh}, with the identification of $t$ with $T_{rel}$ and $\mathcal{E}$ with $m$.
What is important for us now is that this final expression does not display any decaying behavior (the $t$ integral naively would lead a divergent answer, but that will be regularized by compactification of $t$). This will be in contrast with the naive path integral result on the double cone.

Now, let's consider directly computing the path integral on the double cone. 
As discussed before, the path integral is given by a non-linear sigma model with the double cone as the target space:
\be
Z_{\text{DC,obs}}=\int {\mathcal{D}x_L\mathcal{D}x_R\over U(1)}e^{- I_L- I_R};~~~-I_{L,R}=\mp\i{m\over 2}\int_0^{\mathcal{T}_{L,R}}\d u\sin^2\theta_{L,R} \, \dot{t}_{L,R}^2 - \dot{\theta}_{L,R}^2 - \cos^2\theta_{L,R} \, \dot{\Omega}_{d-1}^2+1
\ee
where $t$ is periodically identified $t\sim t+T$, and the $U(1)$ quotient represents the residual U(1) isometry of the double cone. As the path integral factorizes, we will focus on only the right side $R$.
The classical saddle is given by a static trajectory at the center of the static patch, this is described by configuration:
\be\label{eqn:saddle}
t={T\over \mathcal{T}}u+t_0,~~~\theta={\pi\over 2}
\ee
The action of perturbation around this classical saddle is given by:
\be\label{eqn:obpi}
-I=-\i {m\over 2}( \mathcal{T}+{ T^2\over \mathcal{T}})-\i {m\over 2}\int_0^{\mathcal{T}} \d u \dot\epsilon^2 +\mathrm{i} \frac{m}{2} \sum_{i=1}^{d} \int_0^{\mathcal{T}} \mathrm{d}u \, \big( \dot{x}_i^2 + {T^2\over \mathcal{T}^2}x_i^2 \big)
\ee
where $\epsilon$ represents the fluctuation of $t$ and \( x_i \) represents the transverse fluctuations of the particle.
The $\epsilon$ integral is straightforward by expanding in fourier modes $\epsilon=t_0+\sqrt{2}\sum_{n\geq 1} a_n \cos{{2\pi n u\over \mathcal{T}}}+b_n \sin{{2\pi n u\over \mathcal{T}}}$, which leads to a zero mode integral:
\be\label{eqn:tloop}
\int \mathcal{D}\epsilon e^{-\i {m\over 2}\int_0^{\mathcal{T}} \d u \dot\epsilon^2}=\int \d t_0\prod_{n\geq 1}\d a_n \d b_n e^{-\i{2\pi^2  m\over \mathcal{T}}\sum_{n\geq 1} n^2 (a_n^2+b_n^2)}=\sqrt{\i  m\over 2\pi\mathcal{T}}\int \d t_0.
\ee
We can think of this $\epsilon$ integral as coming from the introduction of the clock system.

The $x_i$ integral is the same as the Lorentzian partition function of the inverted harmonic oscillators, which can be computed by analytically continuing the partition function of standard harmonic oscillators. The analytic continuation proceeds as follows. We can first rescale $u\rightarrow {\mathcal T\over T} u$ to move the relative ${T^2\over \mathcal{T}^2}$ coefficient in the mass term outside the integral.  Then we can do a contour rotation: \( x_i \to e^{\mathrm{i} \frac{\pi}{4}} x_i \sqrt{{\mathcal{T}\over m T}}\). This contour rotation is chosen to maintain convergence throughout the deformation and does not introduce extra phases, as it is an ultra-local field redefinition. After the contour rotation, the path integral reduces to the Euclidean path integral of ordinary harmonic oscillators with an ``inverse temperature" \( T \).
This leads to an exponential suppression factor for each transverse direction:
\be\label{eqn:transloop}
\int \mathcal{D}x_i e^{-\int_0^T\dot x_i^2+x_i^2}={1\over 2\sinh {T\over 2}}.
\ee
Since there are $d$ transverse directions, this leads to an overall $e^{-{d\over 2} T}$ decaying behavior. Of course, this factor is not unfamiliar, it just represents the quasi-normal mode decay of heavy particle in de Sitter space that we have discussed before, for example see equation \eqref{eqn:QNM}. Since the universe is expanding, the quantum fluctuations of the particle result in an exponential deviation from the classical trajectory controlled by the classical Lyapunov exponent. Notice also that the Lorentzian partition function does not contain any nontrivial phases, in contrast with the Euclidean path integral of the inverted harmonic oscillators \cite{Maldacena:2024spf,Ivo:2025yek}.

The discussion about the path integral of the left particle is the same as above, we also left with a zero mode parametrize the initial time of the left particle. On double cone, one of these two zero mode is canceled by the U(1) quotient, and we are left with only the relative time shift, which integrates to $T$.  Putting them together, this leads to the partition function on a double cone with fixed periodicity $T$:
\be
Z_{\text{DC,obs}}(\mathcal{T}_L,\mathcal{T}_R)= {m T \over 2\pi \sqrt{\mathcal{T}_L\mathcal{T}_R}} e^{\i {m\over 2}(\mathcal{T}_L-\mathcal{T}_R+{T^2\over \mathcal{T}_L}-{T^2\over \mathcal{T}_R})} ({1\over 2\sinh{T\over 2}})^{2d}.
\ee
To get the full partition function, we should also integrate over double cones with different periodicities as in \eqref{eqn:DCSum}. In other words, the full double cone partition function should be:
\be\label{eqn:dcint}
Z_{\text{DC}}(\mathcal{T}_L,\mathcal{T}_R)=\int \d T \mu(T) Z_{\text{DC,obs}}(\mathcal{T}_L,\mathcal{T}_R)=\int \d T {\mu(T)\over (2\sinh{T\over 2})^{2d}} T {m \over 2\pi \sqrt{\mathcal{T}_L\mathcal{T}_R}} e^{\i {m\over 2}(\mathcal{T}_L-\mathcal{T}_R+{T^2\over \mathcal{T}_L}-{T^2\over \mathcal{T}_R})} .
\ee
where we introduced $\mu(T)$ as the density of double cones of periodicity $T$.
This above formula is the analog of the double trumpet integral in JT gravity, which takes the form (see equation (134) in \cite{Saad:2019lba}):
\be\label{eqn:JTtrumpet}
Z_{0,2}(\beta_1,\beta_2)=\int_0^{\infty} b\d b \left({\gamma^{1/2}\over (2\pi)^{1/2} \beta_1^{1/2}} e^{-{\gamma b^2\over 2\beta_1}}\right)\left({\gamma^{1/2}\over (2\pi)^{1/2} \beta_2^{1/2}} e^{-{\gamma b^2\over 2\beta_2}}\right).
\ee
After identification of $b\rightarrow T,\gamma\rightarrow m,  \beta_{1,2}\rightarrow \mp\i \mathcal{T}_{R,L}$, we see that there are two differences in the above two formulas. The first is the $\i {m\over 2}(\mathcal{T}_L-\mathcal{T}_R)$ piece, which is simply a shift of ground state energy. The second is the additional ${\mu(T)\over (2\sinh{T\over 2})^{2d}}$ term, as mentioned, the $({1\over 2\sinh{T\over 2}})^{2d}$ piece represents the quasi-normal mode decay of the two observers, and $\mu(T)$ is an undetermined measure factor of the double cone. Our next goal is to argue these two factors cancel each other.
One clue that these two factors should cancel each other comes from expression \eqref{eqn:dctr}, which can be related to \eqref{eqn:dcint} via integral transformation of $\mathcal{T}_R-\mathcal{T}_L$ to go to microcanonical ensemble:
\be\label{eqn:canonicalresult}
\int \d \lambda g(\lambda) Z_{\text{DC}}(\mathcal{T}-{\lambda\over 2},\mathcal{T}+{\lambda\over 2})={ \mathcal{T}\over 2\pi}\int \d \left({mT^2\over  2\mathcal{T}^2}\right) f({mT^2\over 2\mathcal{T}^2}-{m\over 2})^2 {\mu(T)\over (2\sinh{T\over 2})^{2d}}.
\ee
where $g(\lambda)$ is the inverse fourier transformation of the window function $f(\mathcal{E})^2$. The factor of $\mathcal{T}$ represents the compactification of the $t$ integral in \eqref{eqn:dctr}, and the rest $\mathcal{E}$ integral already suggests us that $\mu(T) \sim e^{d T}$.

To properly determine the measure factor \( \mu(T) \), we must calculate the density of double cones as a function of \( T \). 
Let's recall that the goal is to sum over geometries with the presence of two observers whose clock winds around $\mathcal{T}$ and $-\mathcal{T}$ respectively. To carry out the path integral, we use an intermediate step by first putting the two observers in an ambient global de Sitter space, and sum over configurations where the beginning and end point of the observers' trajectory is related by a boost symmetry. After quotient of the boost symmetry, we obtain the double cone geometry with the desired property of the observers that the clock winds around $\mathcal{T}$ and $-\mathcal{T}$, as well as compactify the relative time shift. However, in the ambient de Sitter space, there are multiple trajectories a falling observer can follow and  that leads to the construction of various double cone geometries. These different configurations have to be quotient out by the de Sitter isometry group, but the result of the quotient may lead to a nontrivial measure.
More precisely, if we denote the boost symmetry that was used to construct the double cone by the group element \( g_T = e^{-\i K T} \) in \( G =SO(D,1)\). Then, every conjugate of \( g_T \) in the conjugacy class \( [g_T] = \{ h g_T h^{-1} \,|\, h \in G \} \) corresponds to a double cone with the same periodicity \( T \). These double cones are related to the original double cone by a de Sitter isometry $h$ in \( SO(D,1) \) and can be interpreted as describing the return probability for observers moving along different trajectories, where their cosmic horizons are related to the original by \( h \). And since the double cone partition function $Z_{\text{DC,obs}}$ depends only on $T$, it is a class function.
Formally, the volume of \( [g_T] \) is infinite, as the orbit generated by conjugation of \( g_T \) in \( G \) is non-compact. This infinite volume will be resolved by gauging out the de Sitter isometry \( G \), which maps different observers back to the original static patch. The final result after gauging will yield the nontrivial measure factor \( \mu(T) \) associated with each conjugacy class labeled by \( T \).
To calculate $\mu(T)$, we need to compute the Jacobian of the change of variables between the integration over $[g_T]$ and the integration over $G$ (or more precisely, its elements that do not commute with $g_T$), which is closely related to the derivation of the measure factor in the Weyl Integration Formula. Schematically, we have the following integral:
\be\label{eqn:Gint}
\int {\d T \d [g_T]\over G} Z_{\text{DC,obs}}=\int \d T \mu(T) Z_{\text{DC,obs}},
\ee
where $\d T \d[g_T]$ represents integration over the family of conjugacy classes $\lbrace[g_T]\,|\, T\in (0,\infty)\rbrace$, and the quotient of $G$ acts adjointly on $g_T$. The integration measure on $\lbrace[g_T]\,|\, T\in (0,\infty)\rbrace$ is naturally induced from the Haar measure of $G$, which is flat with respect to the Maurer–Cartan form $\omega=g^{-1}\d g$.
Evaluated at $g= h g_T h^{-1}$, this becomes:
\be
\omega=h(g_T^{-1}\d \mathfrak{h} g_T-\d \mathfrak{h}-\i K \d T)h^{-1}.
\ee
 Here, \( -\i K \, \d T = g_T^{-1} \, \d g_T \) represents the variation of \( g_T \), and \( \d \mathfrak{h} = h^{-1} \, \d h \) represents the variation of \( h \). Since the Haar measure is left- and right-invariant, we can ignore the overall conjugation by \( h \), leading to the final decomposition:
\be\label{eqn:MCform}
\omega = (\text{Ad}_{g_T^{-1}} - 1) \, \d \mathfrak{h} - \i K \, \d T.
\ee
We observe that \( \omega \) vanishes for the group generators that commute with \( g_T \); these correspond to the residual \( SO(d) \) symmetry (as well as the \( U(1) \) symmetry) of the double cone, which rotates the transverse sphere. Acting with such symmetries does not generate new double cones, so we will simply throw them away.
The integral over the remaining generators will be fixed by gauging out \( G \), which has a flat measure with respect to \( \d \mathfrak{h} \). 
Comparing with \eqref{eqn:MCform}, we find that after gauging, we are left with a non-trivial Jacobian factor given by:
\be
\mu(T)=|\det \,' (\text{Ad}_{g_T^{-1}} - 1)|,
\ee
where the prime indicates that the determinant is evaluated over the generators that do not commute with \( g_T \).
So we are left with evaluating the determinant. In our case, we have $2d$ generator do not commute with $K$, and they form $d$ $SL(2)$ algebras. For each of the $SL(2)$, the algebra is given by (using standard notation of $SL(2)$ algebra):
\be
[K,H]=\i P,~~~ [K,P]=\i H,~~~[H,P]=\i K.
\ee
After exponential $K$, we find:
\be\label{eqn:mut}
\mu(T)=|\det\,' (\text{Ad}_{g_T^{-1}}-1)|=(2\sinh{T\over 2})^{2d},
\ee
that cancels the one-loop factor $({1\over 2\sinh{T\over 2}})^{2d}$ precisely. 
This leads to our final expression for the observer's partition function on double cone:
\be
Z_{\text{DC}}(\mathcal{T}_L,\mathcal{T}_R)=\int_0^{\infty} \d T\, T {m \over 2\pi \sqrt{\mathcal{T}_L\mathcal{T}_R}} e^{\i {m\over 2}(\mathcal{T}_L-\mathcal{T}_B+{T^2\over \mathcal{T}_L}-{T^2\over \mathcal{T}_R})},
\ee
or after going to the microcanonical ensemble:
\be\label{eqn:Zdc}
Z_{\text{DC}}={\mathcal{T}\over 2\pi}\int \d \mathcal{E} f(\mathcal{E})^2.
\ee

If we include the matter partition function from \eqref{eqn:QFTtr}, the final expression for the double cone partition function in de Sitter spac becomes:
\begin{equation}\label{eqn:ZDC}
Z_{\text{DC}} = \Tr_{L^2(\mathbb{S}^1)} \left[ f(\mathcal{E})^2 \right]\Tr_{\text{QFT}} e^{-i \tilde{K}_m \mathcal{T}}  = \left( \frac{\mathcal{T}}{2\pi} \int \mathrm{d}\mathcal{E} \, f(\mathcal{E})^2 \right) \times \Tr_{\text{QFT}} e^{-i \tilde{K}_m \mathcal{T}}.
\end{equation}
As in the black hole case \cite{Chen:2023hra}, this expression can be interpreted as a modified trace over the bulk Hilbert space $L^2(\mathbb{R}_+) \otimes \tilde{\mathcal{H}}_{\text{QFT}}$: For the observer sector $L^2(\mathbb{R}_+)$, the modification arises from compactifying the $t$-integral in \eqref{eqn:dctr} to the interval $\mathcal{T}$, which defines the meaning of $\Tr_{L^2(\mathbb{S}^1)}$. For the matter sector, the trace involves evolution by the modified boost operator $\tilde{K}_m$, which exhibits quasi-normal mode decay at late times.
Consequently, after a few Hubble times, the double cone contribution leads to linear ramp growth in the SFF, consistent with random matrix theory predictions.

We conclude this section with a few remarks:
\begin{itemize}
    \item We emphasize that the use of the Haar measure for $G$ in \eqref{eqn:Gint} is crucial for obtaining the measure factor $\mu(T)$. In principle, one could attempt to derive $\mu(T)$ directly from the ultralocal measure in gravity, which would be an interesting subject. Instead, we have provided a plausible argument for $\mu(T)$, supported by the observation that the canonical quantization formalism alone do not display any exponential decay term in \eqref{eqn:canonicalresult}---a feature that must therefore be accounted for by $\mu(T)$. Physically, we are saying that after gauge fixing, the time evolution of the observer system is always unitary, and therefore it cannot exhibit any decaying behavior. The exponential decay of the one-loop piece is a common feature in semiclassical chaotic systems, it is a consequence of the classical Lyapunov exponent\footnote{Here we are talking about the expansion of the classical trajectory of the observer. Do not confuse this with the quantum Lyapunov exponent associated with the de Sitter horizon.} that represents the instability of the classical trajectory \cite{haake1991quantum}. In this analog,\footnote{We thank Douglas Stanford for suggesting us this analogy.} $\mu(T)$ plays a role similar to the density of periodic orbits in the Hannay-Ozorio de Almeida sum rule from periodic orbit theory, where it cancels the one-loop determinant in Berry's diagonal approximation for SFF. In that context, the SFF can be expressed as summation of periodic orbits $\gamma,\gamma'$ with period $T$:
    \be
    \text{SFF}\sim {1\over 2\pi} \sum_{\gamma,\gamma'} A_{\gamma} A_{\gamma'}^* e^{\i S_{\gamma}-\i S_{\gamma'}}\approx {T\over 2\pi} n(T)|A_{T}|^2\sim {T\over 2\pi}.
    \ee
   The second approximation is Berry's diagonal approximation, which restricts the sum to pairs \( \gamma = \gamma' \) (up to a time shift). The linear \( T \)-dependence arises from integrating over relative time shifts. As mentioned, one-loop correction \( |A_T|^2 \) decays as \( e^{-\lambda T} \), with \( \lambda \) being the classical Lyapunov exponent, and the \( n(T) \) represents the density of orbits with a marked point, which grows as \( e^{\lambda T} \). The exact cancellation between \( n(T) \) and \( |A_T|^2 \) is essential for recovering the ramp behavior.

\item It is interesting to compare with AdS JT gravity, which is UV-complete and equipped with a well-defined symplectic measure. In that context, the symplectic measure is fixed either by BF theory or the ultralocal metric measure (for the sphere topology) \cite{Saad:2019lba}, and no analogous $\mu(T)$ factor appears---see \eqref{eqn:JTtrumpet}. The discrepancy stems from the choice of boundary particle measures: while JT gravity employs the symplectic measure for the Schwarzian particle \cite{Stanford:2017thb}, our analysis relies on the ultralocal measure. The difference between these measures precisely cancels the one-loop contribution, explaining the absence of $\mu(T)$ in the JT case.

Alternatively, we can consider computing the Schwarzian partition function with the ultralocal measure instead. A one-loop factor of $\frac{1}{\sinh \frac{b}{2}}$ would emerge (see Appendix \ref{app:Sch}), corresponding to the quasi-normal decay of a charged particle in AdS$_2$, as discussed around \eqref{eqn:transloop}. In this scenario, our logic similarly predicts an additional measure factor $\mu(b)$ when gluing two trumpets. 

The essence of our argument is as follows: Our primary objective is to compute the path integral over AdS geometries connecting two Schwarzian particles. To approach this, we examine an auxiliary system consisting of two charged particles in AdS$_2$ with the ultralocal measure, whose propagator endpoints are related through a boost evolution $g_T = e^{-i K T}$ in $G = PSL(2,\mathbb{R})$ that represents a bulk trace. 
This auxiliary system connects to our original problem through a gauging procedure that mods out the $PSL(2,\mathbb{R})$ isometry, which relates different boost evolutions through conjugations.
When evaluating the path integral using the Haar measure for $g_T$, this $PSL(2,\mathbb{R})$ symmetry leads to a divergent volume factor. In the physical configuration space where we gauge fix this isometry, the divergence is removed upon quotienting by $G$, and the gauge-fixing procedure produces the finite measure factor $\mu(T)$. 

A further evidence is supported by BF theory, where the ultralocal measure for boundary particles indeed prescribes the trumpet gluing measure as the Haar measure of the holonomy (see equation (155) in \cite{Kapec:2019ecr}). 

It would be very interesting to investigate whether an analogous calculation using the symplectic measure exists in our de Sitter context. A natural starting point is the sphere path integral, where the configuration space is the loop space in $S^D=SO(D+1)/SO(D)$, modulo the isometry group $SO(D+1)$.

\item In \eqref{eqn:saddle}, we have considered only the winding one configuration. It is interesting to consider also the higher winding solutions on the double cone, that would correspond to the existence of multiple observers.\footnote{We thank Yiming Chen for asking this question.} In general, the transverse fluctuation \eqref{eqn:transloop} of a winding $n$ configuration would lead to $e^{-{d\over 2} n T}$ decaying behavior, that corresponds to the quasi-normal decay of $n$ observers. Only a single observer can be gauge fixed to the center of the static patch, that corresponds to the cancellation with the double cone measure $\mu(T)$, and the rest of the observers will eventually decay into the de Sitter horizon. Therefore at late time only the $n=1$ configuration contributes.

\item In the above discussion, it is important that we construct the double cone geometry using the vaccum de Sitter space which has the full de Sitter isometry. If instead we start with the Schwarzchild-de Sitter space, the conclusion will be different. There, the observer will still have exponential decaying one-loop effect govern by the quasi-normal frequencies on the Schwarzchild-de Sitter geometry. But the analysis of the measure factor $\mu(T)$ will be different, as the Schwarzchild-de Sitter geometry only preserves a $U(1)\times SO(d)$ subgroup, and so there is no nontrivial measure factor comes from gauge fixing. As a result, the matter partition function should decay on such wormhole geometry. Another way to understand the same quasi-normal mode decay is when the black hole is small. In such case, we can view it as ordinary massive matter excitations, and therefore it should also decay into the static patch horizon associated to the observer, as in \eqref{eqn:ZmQNM}.

\item In the above discussion, we argue that in the presence of an observer, the Schwarzchild-de Sitter double cone does not contribute at late time. One might wonder if we can use the black hole as an observer and consider just the Schwarzchild-de Sitter double cone without the observer.\footnote{We thank Yiming Chen and Douglas Stanford for raising this question.} Of course, in general the black hole horizon and de Sitter horizon will not have the same temperature and one has to consider special black holes, such as Nariai black hole or near extremal charged black holes, in order to maintain thermal equilibrium. In such case, the double cone will be a torus, and there seems to be a physical twist mode corresponds to the torus moduli. Nevertheless, in the presence of bulk matter fields, no boost-invariant states exist for general twist configurations (the only boost invariant state is the Hartle-Hawking state prepared by Euclidean path integral on $S^2\times S^{D-2}$) and therefore the Schwarzchild-de Sitter double cone actually does {\it not} lead to a linear in $\mathcal{T}$ growth. The underlying reason is that in the presence of matter field, the Schwarzchild black hole is coupled with the de Sitter horizon due to Hawking radiation, therefore the Hamiltonian constraint does not factorize into sum of $H_{\text{BH}}+H_{\text{dS}}$. In particular, the black hole energy $H_{\text{BH}}$ will not commute with the Hamiltonian constraint to generate the nontrivial $L^2(\mathbb{R}_+)$ clock Hilbert space that was used in the double cone partition function \eqref{eqn:ZDC}.
For our observer model, we have ignored the potential interaction between the observer and bulk matter fields, which seems to be a reasonable approximation when the matter fields are in the Hartle-Hawking state. 
This allows us to use the clock energy $\mathcal{E}$ to generate the  $L^2(\mathbb{R}_+)$ clock Hilbert space.

\item So far, we have focused on the conjugacy class of the boost element $g_T$, but more general conjugacy classes exist in the group $G$, such as those involving additional rotations on the spatial sphere $S_{d-1}$. In general, the conjugacy classes of a Lie group $G$ are in one-to-one correspondence with its maximal torus $T$, modulo the discrete Weyl symmetry. Geometries constructed by quotienting with respect to elements in $T$ can be interpreted as contributing to the spectral form factor with chemical potentials turned on.

We expect such geometries to become relevant when the observer carries charge under the $SO(d)$ rotation group—for instance, by coupling an additional rotor to the particle. We leave a detailed investigation of this direction to future work.

\item In addition to the continuous global symmetries, such as the $SO(d)$ rotations discussed above, the system also possesses discrete symmetries such as the time-reversal symmetry. The discrete symmetry plays a particularly important role, especially in the context of random matrix theory, where it governs the classification of universality classes. In the case of the double cone geometry, a natural implementation of time-reversal symmetry is to act with time reversal on one of the two observers—say, reversing the time orientation for the left observer relative to the right observer. This operation effectively doubles the contribution, leading to an overall multiplicity factor of two. The specific universality class—for instance, GOE or GSE—should depend on additional factors, such as the particular type of bulk topological theory introduced \cite{Stanford:2019vob}.

\item We have seen that the de Sitter double cone leads to a ramp behavior in the spectral form factor, consistent with the predictions of random matrix theory. A crucial ingredient in the derivation is the introduction of the clock system, which leads to the $L^2(\mathbb{R}_+)$ part of the bulk Hilbert space. This naturally raises the question: what is the microscopic origin of the double cone geometry in a de Sitter context?

At present, we do not know the answer. However, let us suppose that the Hilbert space of the full system in the static patch takes the form as in \eqref{eqn:Hconint}
\begin{equation}
\mathcal{H} = \frac{\mathcal{H}_{\text{dS}} \otimes \mathcal{H}_{\text{clock}}}{\mathbf{H}=0},
\end{equation}
where the total Hamiltonian is constrained to vanish. For simplicity, we have ignored the particle degrees of freedom of the observer and focused only on the clock system. If we make the further assumption that the dynamics of $\mathcal{H}_{\text{dS}}$ are governed by a random matrix, an assumption justified in the regime of late-time dynamics, then the system may be modeled by a random matrix $H_{\text{dS}}$ union a clock of energy $\mathcal{E}$, subject to the constraint
\begin{equation}
H_{\text{dS}} + \mathcal{E} = 0.
\end{equation}

The Lorentzian partition function in this framework takes the form 
\begin{equation}\label{eqn:rmtSFF}
{\Tr_{\mathcal{H}} \,\left[ e^{i \mathcal{E} \mathcal{T}} f(\mathcal{E})\right]\over e^{S_{\text{clock}}}} = \int d\mathcal{E}\d E \rho_{\text{dS}}(E) \delta(E + \mathcal{E}) \, e^{i \mathcal{E} \mathcal{T}} f(\mathcal{E}) = \int \d E \rho_{\text{dS}}(E) e^{-i E \mathcal{T}} f(-E),
\end{equation}
where $\{E,\rho_{\text{dS}}(E)\}$ are the energy and spectral density  of $H_{\text{dS}}$. We thus recover the ordinary partition function of $H_{\text{dS}}$, unmodified by any constraint. Note that in general, when the spectrum of both the clock and $H_{\text{dS}}$ are discrete, there will be no solution to Hamiltonian constraint. In the above expression, we have used the smooth approximation of the spectral density $\rho_{\text{dS}}(E)$ while the clock spectrum could be discrete. This is because the double cone geometry arises from the spectral rigidity of $H_{\text{dS}}$ that is not sensitive to the discreteness of the spectrum. It is tempting to speculate the story going beyond this region, to energy level distance of order $e^{-S_{\text{dS}}}$, then in order to still satisfy the Hamiltonian constraint for a finite entropy clock, we either need fine tuned corrections of the Hamiltonian constraint, or $H_{\text{dS}}$ could not be described by a single theory, but instead an ensemble of theories.

Finally, given that de Sitter space admits an $SO(d)$ isometry group, it is natural to expect that $H_{\text{dS}}$ should exhibit a corresponding global $SO(d)$ symmetry, as well as discrete symmetries such as time reversal. These structures could be probed by introducing additional features to the observer system that carries nontrivial representation under $SO(d)$. A more complete understanding of this microscopic picture remains an open and intriguing direction.

\end{itemize}

\section{Discussion}
We have argued in this letter that the double cone wormhole geometry originally constructed by Saad, Shenker, and Stanford admits a natural generalization to the static patch of de Sitter space. Our analysis proceeds via the gravitational path integral, where the inclusion of an observer—and, in particular, an associated clock system—plays a central role. The emergence of the ramp behavior is tied to the relative time shift between the two clocks located in the opposite static patches. 

An additional subtlety concerns the quasi-normal mode decay of the observer. As we explained in Section~\ref{sec:obsdc}, the associated one-loop contribution should be canceled by a measure factor $\mu(T)$ associated with the double cone moduli space.

Along the way, we constructed the observer’s Hilbert space after gauging the global de Sitter isometries in dS$_D$, as discussed in Section~\ref{sec:Obs}. There we showed that the two-sided Hilbert space is naturally given by $L^2(\mathbb{R}_+)$, generalizing the construction in \cite{Chandrasekaran:2022cip}. By matching with sphere partition function, we derive the Hartle-Hawking wavefunction, and use it to demonstrate that the disconnected Lorentzian partition function exhibits exponential decay at late times (see equation~\eqref{eqn:Zltime}).

As already discussed in \cite{Chen:2023hra}, the matter partition function on the double cone can be understood as the trace of an evolution operator generated by a modified boost operator, $\Tr[e^{-i \tilde{K}_m T}]$. The spectrum of $\tilde{K}_m$ is determined by the quasi-normal frequencies, and as a result, the matter partition function decays rapidly—approaching a constant of order one after a few Hubble times, which also suggests that the Thouless time of the system is equal to the Hubble time. 

Finally, around equation~\eqref{eqn:rmtSFF} we briefly discussed a simple random matrix perspective of understanding the emergence of the double cone geometry upon inclusion of the clock system.

Several natural questions remain open:
\begin{itemize}
       \item As we mentioned in the introduction, our microscopic interpretation relies on the assumption that the mysterious minus sign in the sphere partition function \cite{Maldacena:2024spf} does not disrupt the state-counting interpretation. It is therefore important to understand the physics of the minus sign, and whether there exists a subtle way to resolve it in the sphere partition function. In \cite{Maldacena:2024spf}, it was suggested that the extra minus sign could be related to the fact that the de sitter horizon area is a local maximum (bounce) rather than a local minimum (throat) as for the black hole horizon \cite{Engelhardt:2023bpv}. In \cite{Ivo:2025yek, Shi:2025amq}, a generalization to $S^p\times M_q$ manifolds was considered and the analysis shows up to extra phases due to physical negative modes, the phase is the same as that of $S^p$. It would be interesting to understand whether such a minus sign arises for more general bounces and, more importantly, what its microscopic interpretation might be.
         \item A further direction is to enrich the observer model so as to incorporate the full $SO(d)$ rotation symmetry, along with possible discrete symmetries of dS$_D$. This may shed light on the symmetry structure of the random matrix ensemble, as discussed at the end of Section~\ref{sec:obsdc}.
       \item An alternative approach is to consider simplified models of de Sitter gravity that isolate the random matrix behavior of the static patch horizon, that is agnostic to spatial locality. The static patch version of de Sitter JT gravity, mentioned in the introduction, provides a natural toy model in this direction. As briefly noted, in that framework the gravitational degrees of freedom reduce to a pair of boundary particles localized at the two static patches. Much of the analysis presented in Section~\ref{sec:Obs} can be adapted to this case, provided one employs the ultra-local measure for the boundary particle as well.

We expect this model to provide a useful testing ground for the ideas discussed here, and we will report further progress in an upcoming work~\cite{HuYangZhangZheng:upcoming}.

    \item If the static patch of de Sitter space, in the presence of an observer, indeed admits a dual quantum mechanical description, we can interpret the wormhole geometries as probes of the underlying random matrix statistics of the Hamiltonian. In this framework, the ramp growth corresponds to spectral rigidity, as mentioned. Building on random matrix universality, we anticipate the existence of a family of more intricate (nearly on-shell) wormhole geometries, governed by higher-order corrections in the spectral correlators (such as those with time reversal symmetry)\cite{Saad:2022kfe}. The observer's trajectory in these geometries would likely correspond to encounter-type configurations, analogous to those in the language of periodic orbits, as if the de Sitter space had finite volume. In this framework, the double cone geometry represents Berry's leading diagonal approximation. For the encounter geometries, we expect the local geometry near the encounters to reflect shockwave configurations, which could be different from those in the black hole case. We leave the detailed exploration of this problem to future investigations. 
    \item If we send a shockwave through de Sitter horizon, the two static patches become causally connected \cite{Aalsma:2020aib}, how shall we understand this? In black hole case, quantum mechanics guarantees that any unitary acting on one side cannot make the wormhole traversable, perhaps one should think of creating shockwave in de Sitter space as a kind of projection operation? 
    \item The emergence of double cone geometries suggests a breakdown of the standard notion of time under sufficiently long evolution. In black hole physics, this phenomenon is tied to the typical-state firewall paradox \cite{Stanford:2022fdt}. Does an analogous de Sitter version of this paradox exist? Here one difference is the lack of singularity behind the de Sitter horizon; the other difference is that the observer may not be able to live long enough.
\end{itemize}

\section*{Acknowledgements} 

We would like to thank Douglas Stanford for raising the question of the one-loop determinant of the observer. We would like to thank Yiming Chen, Steve Shenker and Douglas Stanford for insightful comments on the draft. We would also like to thank Frank Ferrari, Xuyao Hu, Jingru Lu, Juan Maldacena, Don Marolf, Manqian Ou, Sirui Shuai and Zimo Sun for helpful discussions. ZY acknowledges support from NSFC Grant No. 12447108, 12342501, 12475071, 12247103.

\appendix
\section{Plancherel measure}\label{app:planch}
The Plancherel measure is determined by the asymptotic behavior of \( \psi_s(t) \) \eqref{eqn:spectrum} at large \( t \):
\begin{equation}
\psi_s(t \to \infty) = c_s e^{- (d/2 + \i s)t} + c_s^* e^{- (d/2 - \i s)t},
\end{equation}
with coefficient \( c_s \):
\begin{equation}
c_s = 2^{d-1} \frac{\Gamma \left( \frac{d+1}{2} \right) \Gamma(-\i s)}{\sqrt{\pi} \Gamma \left( \frac{d}{2} - \i s \right)}.
\end{equation}
From this, the integral at large \( t \) takes the form
\begin{equation}
\int \d t \, 2^{-d} c_s^* c_{s'} e^{\i (s' - s)t} \approx \frac{\delta(s - s')}{\rho(s)},
\end{equation}
which determines the Plancherel measure as
\begin{equation}
\rho(s) = \frac{2^d}{2\pi |c_s|^2}.
\end{equation}

\section{Scalar partition function in dS$_2$ double cone}\label{app:scalardS2}
In this appendix, we will perform an one-loop analysis of the double cone partition function of a scalar field $\phi$ in two dimensions, with action:
\be
-I=-\int \partial_{\mu}\phi \partial^{\mu}\phi+m^2\phi^2.
\ee
The metric is given by:
\be
\d s^2=-\sin^2(\theta-\i \epsilon)\d t^2+\d \theta^2,\quad \theta \in [-{\pi\over 2},{\pi\over 2}]
\ee
We will consider only half of dS$_2$ and impose the dirichlet boundary condition that $\phi=0$ at the two boundaries $\theta=\pm{\pi\over 2}$.

To compute the one-loop determinant, we want to expand the field $\phi$ in terms of eigenfunctions of the Laplacian operator:
\be
\nabla^2=-{1\over \sin^2(\theta-\i \epsilon)}\partial_t^2+{1\over \sin (\theta-\i \epsilon)}\partial_{\theta} \sin(\theta-\i \epsilon) \partial_{\theta}
\ee
which naively does not satisfy the Sturm-Liouville theorem due to the fact that $\sin(\theta-\i \epsilon)$ is not always positive. In order to get around that problem, let's consider a change of variables from $\theta$ to $\eta$ such that $\sin( \theta)={1\over \cosh\eta}$. In this new coordinate, the metric now becomes:
\be
\d s^2={1\over \cosh^2\eta}(-\d t^2+\d \eta^2),
\ee
with $\eta$ goes from $0$ to $\i \pi$ with a contour that the real part of $\eta$ is positive, which can be further deformed to be parallel to the imaginary axis so that $\eta=\i x$. This leads to a negative EAdS$_2$ metric (but with different range of the radial direction):
\be\label{eqn:EAdS}
\d s^2=-{1\over \cos^2 (x-\i \epsilon)}(\d t^2+\d x^2),\quad x\in[0,\pi].
\ee
The action in this new set of coordinates becomes:
\be
-I=\int \d t \d x \phi(-\partial_t^2-\partial_x^2-{m^2\over \cos^2 x})\phi.
\ee
Now the operator $\partial_x^2+{m^2\over \cos^2x}$ with vanishing boundary condition as $x=0$ and $\pi$ indeed satisfies the Sturm-Liouville theorem (at least for small enough $m^2$), and therefore its eigenfunctions form a complete basis of $L^2$ function of $x$. Specifically, those are:
\bea
(\partial_x^2+{m^2\over \cos^2x}) f_{n,\pm}(x)=-\lambda_{\pm}(n) f_{n,\pm}(x),\quad \lambda_{\pm}(n)=({1\over 2}+1+2n\pm \sqrt{1/4-m^2})^2,\quad n\geq 0.\\
f_{n,\pm}(x)=\cos(x)^{{1\over 2}\pm \sqrt{1/4-m^2}} {}_2F_1(-{1\over 2}-n,1+n\pm \sqrt{1/4-m^2},1\pm \sqrt{1/4-m^2},\cos^2x).
\eea
We can then expand $\phi$ in the $f_{n,\pm}$ basis, that gives:
\be
\phi(t,x)=\sum_{n,\pm} f_{n,\pm} (x) \phi_{n,\pm}(t).
\ee
Plug in the action, this leads to a sum of harmonic oscillators with mass squared $\lambda_{\pm}(n)$:
\be
-I=\sum_{n,\pm}\int \d t \phi_{n,\pm}(t)(-\partial_t^2+\lambda_{\pm}(n))\phi_{n,\pm}(t),
\ee
Let's consider the situation of $m^2$ being negative, and the case of positive $m^2$ can be obtained through analytic continuation. We note that in such case $\lambda_{\pm}(n))$ are all positive definite. After analytic continue $\phi\rightarrow \i \phi$, the path integral is convergent, and the one-loop determinant is given by product of the partition function of harmonic oscillators without any overall phases:
\be
\prod_{n,\pm}{1\over 2\sinh{\sqrt{\lambda_{\pm}(n)}T\over 2}}=e^{\# T}\prod_{\omega} {1\over 1-e^{-\i \omega T}},\quad \omega=\i |{1\over 2}+1+2n\pm \sqrt{1/4-m^2}|.
\ee
The overall ground state energy should be canceled by the proper UV regularization using the metric \eqref{eqn:EAdS}, which will be an interesting thing to work out explicitly. Here we will just declare the ground state energy has to vanish due to the fact that the Hartle-Hawking state is explicit invariant under the boost evolution. After taking that into account, we recovers the result \eqref{eqn:ZmQNM} in our special case.

\section{Lorentzian partition function of harmonic oscillators and inverted harmonic oscillators}
It is somewhat interesting that the double cone partition function of negative mass squared field does not seem to contain any nontrivial phases. In this section, we examine this phenomenon for simple inverted harmonic oscillators, and find in contrast to the Euclidean partition function, the Lorentzian partition function of inverted harmonic oscillator does not receive any phases.

Let's start with the Lorentzian partition function of normal harmonic oscillator ($\omega>0$):
\be
Z(T)= \int \mathcal{D} x e^{i \int_0^T \d t (\dot x^2-\omega^2 x^2)}.
\ee
We can expand $x(t)$ in fourier basis $x(t)=x_0+\sqrt{2}\sum_{n=1}^{\infty}a_n\cos {2\pi n\over T} t+b_n\sin {2\pi n\over T}  t $, that gives:
\be
Z(T)=\int da_0\prod_{n\geq 1}d a_n db_n e^{-i \omega^2 a_0^2+i \sum_{n\geq 1} (a_n^2+b_n^2)(({2\pi n\over T})^2-\omega^2) }.
\ee
Notice that depends on whether $({2\pi n\over T})^2-\omega^2 $ is big or smaller than zero, the $a_n,b_n$ either has $i$ or $-i$ factor in front of the action, this means as we increase $T$ such that $T$ cross ${2\pi n\over \omega}$ the partition function flips a sign. 

The one-loop integral is straightforward to do, we can slightly deform the integration variables with positive or negative imaginary part. The partition function becomes:
\be
Z(T)={\pm i\over \omega}\prod_{n\geq 1} {1\over ({2\pi n\over T})^2-\omega^2}=\pm i \prod_{n\geq 1}{T^2\over (2\pi n)^2}\prod_{n\geq 1}{1\over 1-({\omega T\over 2\pi n})^2}.
\ee
Using
\be
\prod_{n\geq 1}{1\over n^2}=e^{2\zeta'(0)}={1\over 2\pi},~~~\prod_{n\geq 1}1-({x\over n})^2={\sin (\pi x)\over \pi x},
\ee
we find the partition function becomes:
\be
Z(T)=\pm i {1\over 2 \sin({\omega T\over 2})}.
\ee
We notice that the sine function has oscillation signs when $T$ cross ${2\pi n\over \omega}$ as we anticipated. The $\pm i$ comes from the infinite product:
\be
{1\over \sqrt{i}}\prod_{n\geq 1}{1\over - i}={1\over \sqrt{-1}}
\ee
To determine it, we can add a small euclidean part for $T\rightarrow T- i \epsilon$, which is equivalent to $\omega\rightarrow \omega-i  \epsilon$. That leads to
\be
{1\over \sqrt{-1}}={\omega \over \sqrt{-\omega^2}}=- i.
\ee
This is consistent with the fact that under analytic continuation $T\rightarrow - i\beta$, the partition function becomes the standard Euclidean partition function of harmonic oscillators
\be
Z(-i \beta)={1\over 2\sinh({\omega\beta\over 2})}.
\ee

The upshot is that the Lorentzian partition function is well defined and gives us an imaginary oscillating result that can be obtained from analytic continuation of the Euclidean partition function.

Now let's consider the Lorentzian partition function of inverted Harmonic oscillators. For that we replace $\omega^2\rightarrow -\mu^2$. The quadratic fluctuation becomes:
\be
Z(T)=\int da_0\prod_{n\geq 1}d a_n db_n e^{i \mu^2 a_0^2+i \sum_{n\geq 1} (a_n^2+b_n^2)(({2\pi n\over T})^2+\mu^2) }
\ee
So that the integration contour only requires a small real imaginary part deformation. The one-loop determinant becomes equal to the Euclidean partition function of normal harmonic oscillators:
\be
Z(T)={1\over \mu} \prod_{n\geq 1} {1\over ({2\pi n\over T})^2+\mu^2}={1\over 2\sinh({\mu T\over 2})}.
\ee
The upshot is that the Lorentzian partition function for inverted harmonic oscillator is real positive, without any phases, in contrast to the Euclidean partition function.

\section{Ultralocal measure in AdS$_2$}\label{app:Sch}
In this appendix, we present the one-loop calculation for the Schwarzian particle on the double cone geometry using the ultralocal measure. In AdS$_2$, the double cone geometry arises from the analytic continuation of the double trumpet. This connection can be understood by examining the inverse Laplace transform of the double trumpet partition function:
\be\label{eqn:dcintgralapp}
{1\over 2\pi \i }\int_{-\i \infty}^{\i \infty} \d \beta Z_{0,2}(\beta+\i T, \beta -\i T) e^{2\beta E}={1\over 2\pi \i }\int_{-\i \infty}^{\i \infty}\d \beta \d b b {\gamma^{1\over 2}\over \sqrt{\beta+\i T}}{\gamma^{1\over 2}\over \sqrt{\beta-\i T}} e^{-{\gamma\over 2}{b^2\over \beta+\i T}-{\gamma\over 2}{b^2\over \beta-\i T}+2\beta E}.
\ee
In the large-$T$ limit, the moduli integral over $b$ is dominated by the saddle point, which corresponds to the double cone geometry:
\be
b=\sqrt{2E\over \gamma} T={2\pi T\over \beta_E}, \quad E={2\pi^2\gamma\over \beta}.
\ee
Thus, the analysis reduces to that of the trumpet partition function.

Let's recall that the trumpet partition function \cite{Saad:2019lba} is given by the following Schwarzian path integral:
\be
Z_{\text{Sch}}^{\text{Trumpet}}(\beta,b)=\int {\d \mu(\tau)\over \text{U(1)}}\exp\left[-{\gamma\over 2}\int_0^{\beta}\left({\tau''^2\over \tau'^2}+\tau'^2\right)\right].
\ee
Here, in the standard treatment, $\mu(\tau)$ is taken to be the symplectic measure that can be determined using the BF formalism. We will be interested in the result of the path integral if we use instead the ultralocal measure, and show that the one-loop contribution leads to an exponentially decaying piece ${1\over \sinh {b\over 2}}$.
Expanding around the classical saddle:
\be
\tau(u)={b\over \beta}(u+\epsilon(u)),~~~\epsilon(u)=\epsilon_0+\sum_{n\neq 0} e^{-{2\pi\over \beta} \i n u}(\epsilon_n^{\text{(R)}}+\i \epsilon_n^{\text{(I)}}),
\ee
with $\epsilon_n^{\text{(R)}}=\epsilon_{-n}^{\text{(R)}}$ and $\epsilon_{n}^{\text{(I)}}=-\epsilon_{-n}^{\text{(I)}
}$, the ultralocal measure for $\epsilon$ can be determined from the metric in $\delta\tau$ space:
\be
|\delta \tau(u)|^2=\int_0^{\beta} \d u \delta \tau(u)^2={2b^2\over \beta}\sum_{n\geq 1}((\epsilon_n^{\text{(R)}})^2+(\epsilon_n^{\text{(I)}})^2)+{b^2\over \beta}\epsilon_0^2.
\ee
This takes the form:
\be\label{eqn:ultraepsilon}
\d \mu(\tau)={b\over \sqrt{\beta}}\d \epsilon_0\prod_{n\geq 1} {2b^2\over \beta}\d \epsilon_n^{\text{(R)}}\d \epsilon_n^{\text{(I)}}.
\ee
Here, for the sake of clarity, we have used the convention of \cite{Saad:2019lba}. The relation with the variables in \eqref{eqn:tloop} is
\be
{b\over \beta}\epsilon_0=t_0,\quad {2b\over \beta} \epsilon_n^{\text{(R)}}=\sqrt{2} a_n,\quad {2b\over \beta} \epsilon_n^{\text{(I)}}=\sqrt{2} b_n.
\ee
Using the ultralocal measure \eqref{eqn:ultraepsilon}, we find that the one-loop contribution to the Schwarzian trumpet partition function becomes:
\bea
Z_{\text{Sch}}^{\text{Trumpet}}(\beta,b)&=&e^{-{\gamma\over 2}{b^2\over \beta}} \sqrt{\beta}\prod_{n\geq 1} {2b^2\over \beta}  \int \d \epsilon_n^{\text{(R)}}\d \epsilon_n^{\text{(I)}} e^{-(2\pi)^4\gamma {(\epsilon_n^{\text{(R)}})^2+(\epsilon_n^{\text{(I)}})^2\over \beta^3}(n^4+{b^2\over (2\pi)^2}n^2)},\\
&=&e^{-{\gamma\over 2}{b^2\over \beta}} \sqrt{\beta}\prod_{n\geq 1} {b^2\beta^2\over (2\pi)^3 \gamma (n^4+{b^2\over (2\pi)^2}n^2)},\\
&=&e^{-{\gamma\over 2}{b^2\over \beta}}  {\gamma^{1/2}\over \sqrt{2\pi} \beta^{1/2}} \times {1\over 2\sinh(b/2)}.
\eea
Here the U(1) quotient cancels with the $\epsilon_0$ integral, which leads to the overall $\sqrt{\beta}$ factor in front of the infinite product. Comparing with the result using the symplectic measure, we find an additional overall factor of ${1\over 2\sinh(b/2)}$ when using the ultralocal measure. We can understand this additional factor by considering the Schwarzian particle as a charged particle in AdS$_2$. It belongs to the principal series of the SL(2) representation, that has complex scaling dimension:
\be
\Delta={1\over 2}+\i s,~~~s\geq 0.
\ee
The real part of $\Delta$ controls the leading decaying behavior of the Schwarizan propagator at late times should become $e^{-{\pi T\over \beta}}$, that corresponds to the large $b$ expansion of the ${1\over 2\sinh(b/2)}$ factor. We also observe that preserving the double trumpet partition function in \eqref{eqn:dcintgralapp} requires an additional measure factor of $(2\sinh\frac{b}{2})^2$ in the gluing rule for two trumpets. This can be understood as coming from gauge fixing the PSL(2,$\mathbb{R}$) symmetry of the Schwarzian particle moving in ambient AdS space as discussed in Section~\ref{sec:obsdc}.

\bibliography{references}

\providecommand{\href}[2]{#2}\begingroup\raggedright\begin{thebibliography}{10}

\bibitem{Dyson:2002pf}
L.~Dyson, M.~Kleban, and L.~Susskind, ``{Disturbing implications of a
  cosmological constant},''
  \href{http://dx.doi.org/10.1088/1126-6708/2002/10/011}{{\em JHEP} {\bfseries
  10} (2002) 011}, \href{http://arxiv.org/abs/hep-th/0208013}{{\ttfamily
  arXiv:hep-th/0208013}}.

\bibitem{Goheer:2002vf}
N.~Goheer, M.~Kleban, and L.~Susskind, ``{The Trouble with de Sitter space},''
  \href{http://dx.doi.org/10.1088/1126-6708/2003/07/056}{{\em JHEP} {\bfseries
  07} (2003) 056}, \href{http://arxiv.org/abs/hep-th/0212209}{{\ttfamily
  arXiv:hep-th/0212209}}.

\bibitem{Saad:2018bqo}
P.~Saad, S.~H. Shenker, and D.~Stanford, ``{A semiclassical ramp in SYK and in
  gravity},'' \href{http://arxiv.org/abs/1806.06840}{{\ttfamily
  arXiv:1806.06840 [hep-th]}}.

\bibitem{Chen:2023hra}
Y.~Chen, V.~Ivo, and J.~Maldacena, ``{Comments on the double cone wormhole},''
  \href{http://dx.doi.org/10.1007/JHEP04(2024)124}{{\em JHEP} {\bfseries 04}
  (2024) 124}, \href{http://arxiv.org/abs/2310.11617}{{\ttfamily
  arXiv:2310.11617 [hep-th]}}.

\bibitem{Witten:2021unn}
E.~Witten, ``{Gravity and the crossed product},''
  \href{http://dx.doi.org/10.1007/JHEP10(2022)008}{{\em JHEP} {\bfseries 10}
  (2022) 008}, \href{http://arxiv.org/abs/2112.12828}{{\ttfamily
  arXiv:2112.12828 [hep-th]}}.

\bibitem{Chandrasekaran:2022eqq}
V.~Chandrasekaran, G.~Penington, and E.~Witten, ``{Large N algebras and
  generalized entropy},'' \href{http://dx.doi.org/10.1007/JHEP04(2023)009}{{\em
  JHEP} {\bfseries 04} (2023) 009},
  \href{http://arxiv.org/abs/2209.10454}{{\ttfamily arXiv:2209.10454
  [hep-th]}}.

\bibitem{Chandrasekaran:2022cip}
V.~Chandrasekaran, R.~Longo, G.~Penington, and E.~Witten, ``{An algebra of
  observables for de Sitter space},''
  \href{http://dx.doi.org/10.1007/JHEP02(2023)082}{{\em JHEP} {\bfseries 02}
  (2023) 082}, \href{http://arxiv.org/abs/2206.10780}{{\ttfamily
  arXiv:2206.10780 [hep-th]}}.

\bibitem{Dong:2018cuv}
X.~Dong, E.~Silverstein, and G.~Torroba, ``{De Sitter Holography and
  Entanglement Entropy},''
  \href{http://dx.doi.org/10.1007/JHEP07(2018)050}{{\em JHEP} {\bfseries 07}
  (2018) 050}, \href{http://arxiv.org/abs/1804.08623}{{\ttfamily
  arXiv:1804.08623 [hep-th]}}.

\bibitem{Silverstein:2022dfj}
E.~Silverstein, ``{Black hole to cosmic horizon microstates in string/M theory:
  timelike boundaries and internal averaging},''
  \href{http://dx.doi.org/10.1007/JHEP05(2023)160}{{\em JHEP} {\bfseries 05}
  (2023) 160}, \href{http://arxiv.org/abs/2212.00588}{{\ttfamily
  arXiv:2212.00588 [hep-th]}}.

\bibitem{Kaplan:2024xyk}
M.~Kaplan, D.~Marolf, X.~Yu, and Y.~Zhao, ``{De Sitter quantum gravity and the
  emergence of local algebras},''
  \href{http://arxiv.org/abs/2410.00111}{{\ttfamily arXiv:2410.00111
  [hep-th]}}.

\bibitem{Kolchmeyer:2024fly}
D.~K. Kolchmeyer and H.~Liu, ``{Chaos and the Emergence of the Cosmological
  Horizon},'' \href{http://arxiv.org/abs/2411.08090}{{\ttfamily
  arXiv:2411.08090 [hep-th]}}.

\bibitem{Tietto:2025oxn}
D.~Tietto and H.~Verlinde, ``{A microscopic model of de Sitter spacetime with
  an observer},'' \href{http://arxiv.org/abs/2502.03869}{{\ttfamily
  arXiv:2502.03869 [hep-th]}}.

\bibitem{Maldacena:2024spf}
J.~Maldacena, ``{Real observers solving imaginary problems},''
  \href{http://arxiv.org/abs/2412.14014}{{\ttfamily arXiv:2412.14014
  [hep-th]}}.

\bibitem{Ivo:2025yek}
V.~Ivo, J.~Maldacena, and Z.~Sun, ``{Physical instabilities and the phase of
  the Euclidean path integral},''
  \href{http://arxiv.org/abs/2504.00920}{{\ttfamily arXiv:2504.00920
  [hep-th]}}.

\bibitem{Shi:2025amq}
X.~Shi and G.~J. Turiaci, ``{The phase of the gravitational path integral},''
  \href{http://arxiv.org/abs/2504.00900}{{\ttfamily arXiv:2504.00900
  [hep-th]}}.

\bibitem{Chen:2025jqm}
Y.~Chen, D.~Stanford, H.~Tang, and Z.~Yang, ``{On the phase of the de Sitter
  density of states},'' \href{http://arxiv.org/abs/2511.01400}{{\ttfamily
  arXiv:2511.01400 [hep-th]}}.

\bibitem{Polyakov:1987ez}
A.~M. Polyakov, ``{Gauge Fields and Strings},''
{\em Contemp. Concepts Phys.} {\bfseries 3} (1987) 1--301.
%%CITATION = 00380,3,1;%%.

\bibitem{hawking2023large}
S.~W. Hawking and G.~F. Ellis, {\em The large scale structure of space-time}.
\newblock Cambridge university press, 2023.

\bibitem{Dong:2020uxp}
X.~Dong, X.-L. Qi, Z.~Shangnan, and Z.~Yang, ``{Effective entropy of quantum
  fields coupled with gravity},''
  \href{http://dx.doi.org/10.1007/JHEP10(2020)052}{{\em JHEP} {\bfseries 10}
  (2020) 052}, \href{http://arxiv.org/abs/2007.02987}{{\ttfamily
  arXiv:2007.02987 [hep-th]}}.

\bibitem{Jackiw:1984je}
R.~Jackiw, ``{Lower Dimensional Gravity},''
\href{http://dx.doi.org/10.1016/0550-3213(85)90448-1}{{\em Nucl. Phys.}
  {\bfseries B252} (1985) 343--356}.
%%CITATION = NUPHA,B252,343;%%.

\bibitem{Teitelboim:1983ux}
C.~Teitelboim, ``{Gravitation and Hamiltonian Structure in Two Space-Time
  Dimensions},''
\href{http://dx.doi.org/10.1016/0370-2693(83)90012-6}{{\em Phys. Lett.}
  {\bfseries B126} (1983) 41--45}.
%%CITATION = PHLTA,B126,41;%%.

\bibitem{Jensen:2016pah}
K.~Jensen, ``{Chaos in AdS$_2$ Holography},''
  \href{http://dx.doi.org/10.1103/PhysRevLett.117.111601}{{\em Phys.\ Rev.\
  Lett.} {\bfseries 117} no.~11, (2016) 111601},
  \href{http://arxiv.org/abs/1605.06098}{{\ttfamily arXiv:1605.06098
  [hep-th]}}.

\bibitem{Maldacena:2016upp}
J.~Maldacena, D.~Stanford, and Z.~Yang, ``{Conformal symmetry and its breaking
  in two dimensional Nearly Anti-de-Sitter space},''
  \href{http://dx.doi.org/10.1093/ptep/ptw124}{{\em PTEP} {\bfseries 2016}
  no.~12, (2016) 12C104}, \href{http://arxiv.org/abs/1606.01857}{{\ttfamily
  arXiv:1606.01857 [hep-th]}}.

\bibitem{Engelsoy:2016xyb}
J.~Engels{\"o}y, T.~G. Mertens, and H.~Verlinde, ``{An investigation of
  AdS$_{2}$ backreaction and holography},''
  \href{http://dx.doi.org/10.1007/JHEP07(2016)139}{{\em JHEP} {\bfseries 07}
  (2016) 139}, \href{http://arxiv.org/abs/1606.03438}{{\ttfamily
  arXiv:1606.03438 [hep-th]}}.

\bibitem{Svesko:2022txo}
A.~Svesko, E.~Verheijden, E.~P. Verlinde, and M.~R. Visser, ``{Quasi-local
  energy and microcanonical entropy in two-dimensional nearly de Sitter
  gravity},'' \href{http://dx.doi.org/10.1007/JHEP08(2022)075}{{\em JHEP}
  {\bfseries 08} (2022) 075}, \href{http://arxiv.org/abs/2203.00700}{{\ttfamily
  arXiv:2203.00700 [hep-th]}}.

\bibitem{Rahman:2022jsf}
A.~A. Rahman, ``{dS JT Gravity and Double-Scaled SYK},''
  \href{http://arxiv.org/abs/2209.09997}{{\ttfamily arXiv:2209.09997
  [hep-th]}}.

\bibitem{HuYangZhangZheng:upcoming}
X.~Hu, Z.~Yang, Y.~Zhang, and W.~Zheng To appear, 2025.

\bibitem{Almheiri:2013hfa}
A.~Almheiri, D.~Marolf, J.~Polchinski, D.~Stanford, and J.~Sully, ``{An
  Apologia for Firewalls},''
  \href{http://dx.doi.org/10.1007/JHEP09(2013)018}{{\em JHEP} {\bfseries 09}
  (2013) 018}, \href{http://arxiv.org/abs/1304.6483}{{\ttfamily arXiv:1304.6483
  [hep-th]}}.

\bibitem{Marolf:2013dba}
D.~Marolf and J.~Polchinski, ``{Gauge/Gravity Duality and the Black Hole
  Interior},'' \href{http://dx.doi.org/10.1103/PhysRevLett.111.171301}{{\em
  Phys. Rev. Lett.} {\bfseries 111} (2013) 171301},
  \href{http://arxiv.org/abs/1307.4706}{{\ttfamily arXiv:1307.4706 [hep-th]}}.

\bibitem{Susskind:2012rm}
L.~Susskind, ``{Singularities, Firewalls, and Complementarity},''
  \href{http://arxiv.org/abs/1208.3445}{{\ttfamily arXiv:1208.3445 [hep-th]}}.

\bibitem{Susskind:2015toa}
L.~Susskind, ``{The Typical-State Paradox: Diagnosing Horizons with
  Complexity},'' \href{http://dx.doi.org/10.1002/prop.201500091}{{\em Fortsch.
  Phys.} {\bfseries 64} (2016) 84--91},
  \href{http://arxiv.org/abs/1507.02287}{{\ttfamily arXiv:1507.02287
  [hep-th]}}.

\bibitem{deBoer:2018ibj}
J.~de~Boer, R.~Van~Breukelen, S.~F. Lokhande, K.~Papadodimas, and E.~Verlinde,
  ``{On the interior geometry of a typical black hole microstate},''
  \href{http://dx.doi.org/10.1007/JHEP05(2019)010}{{\em JHEP} {\bfseries 05}
  (2019) 010}, \href{http://arxiv.org/abs/1804.10580}{{\ttfamily
  arXiv:1804.10580 [hep-th]}}.

\bibitem{DeBoer:2019yoe}
J.~De~Boer, R.~Van~Breukelen, S.~F. Lokhande, K.~Papadodimas, and E.~Verlinde,
  ``{Probing typical black hole microstates},''
  \href{http://dx.doi.org/10.1007/JHEP01(2020)062}{{\em JHEP} {\bfseries 01}
  (2020) 062}, \href{http://arxiv.org/abs/1901.08527}{{\ttfamily
  arXiv:1901.08527 [hep-th]}}.

\bibitem{Susskind:2020wwe}
L.~Susskind, ``{Black Holes at Exp-time},''
  \href{http://arxiv.org/abs/2006.01280}{{\ttfamily arXiv:2006.01280
  [hep-th]}}.

\bibitem{Stanford:2022fdt}
D.~Stanford and Z.~Yang, ``{Firewalls from wormholes},''
  \href{http://arxiv.org/abs/2208.01625}{{\ttfamily arXiv:2208.01625
  [hep-th]}}.

\bibitem{Blommaert:2024ftn}
A.~Blommaert, C.-H. Chen, and Y.~Nomura, ``{Firewalls at exponentially late
  times},'' \href{http://dx.doi.org/10.1007/JHEP10(2024)131}{{\em JHEP}
  {\bfseries 10} (2024) 131}, \href{http://arxiv.org/abs/2403.07049}{{\ttfamily
  arXiv:2403.07049 [hep-th]}}.

\bibitem{Iliesiu:2024cnh}
L.~V. Iliesiu, A.~Levine, H.~W. Lin, H.~Maxfield, and M.~Mezei, ``{On the
  non-perturbative bulk Hilbert space of JT gravity},''
  \href{http://dx.doi.org/10.1007/JHEP10(2024)220}{{\em JHEP} {\bfseries 10}
  (2024) 220}, \href{http://arxiv.org/abs/2403.08696}{{\ttfamily
  arXiv:2403.08696 [hep-th]}}.

\bibitem{Yang:upcoming}
Z.~Yang, ``Comments on the saad wormhole.''
  https://groups.oist.jp/exu-oist/recorded-talks, 2024.

\bibitem{ShaghoulianTalks}
E.~Shaghoulian, ``The central dogma and horizons in quantum cosmology
  \href{https://cds.cern.ch/record/2871349}.''.

\bibitem{Susskind:2021omt}
L.~Susskind, ``{De Sitter Holography: Fluctuations, Anomalous Symmetry, and
  Wormholes},'' \href{http://dx.doi.org/10.3390/universe7120464}{{\em Universe}
  {\bfseries 7} no.~12, (2021) 464},
  \href{http://arxiv.org/abs/2106.03964}{{\ttfamily arXiv:2106.03964
  [hep-th]}}.

\bibitem{Kitaev:2018wpr}
A.~Kitaev and S.~J. Suh, ``{Statistical mechanics of a two-dimensional black
  hole},'' \href{http://dx.doi.org/10.1007/JHEP05(2019)198}{{\em JHEP}
  {\bfseries 05} (2019) 198},
\href{http://arxiv.org/abs/1808.07032}{{\ttfamily arXiv:1808.07032 [hep-th]}}.
%%CITATION = ARXIV:1808.07032;%%.

\bibitem{Yang:2018gdb}
Z.~Yang, ``{The Quantum Gravity Dynamics of Near Extremal Black Holes},''
  \href{http://dx.doi.org/10.1007/JHEP05(2019)205}{{\em JHEP} {\bfseries 05}
  (2019) 205},
\href{http://arxiv.org/abs/1809.08647}{{\ttfamily arXiv:1809.08647 [hep-th]}}.
%%CITATION = ARXIV:1809.08647;%%.

\bibitem{Penington:2023dql}
G.~Penington and E.~Witten, ``{Algebras and States in JT Gravity},''
  \href{http://arxiv.org/abs/2301.07257}{{\ttfamily arXiv:2301.07257
  [hep-th]}}.

\bibitem{Jafferis:2019wkd}
D.~L. Jafferis and D.~K. Kolchmeyer, ``{Entanglement Entropy in
  Jackiw-Teitelboim Gravity},''
\href{http://arxiv.org/abs/1911.10663}{{\ttfamily arXiv:1911.10663 [hep-th]}}.
%%CITATION = ARXIV:1911.10663;%%.

\bibitem{helgason2022groups}
S.~Helgason, {\em Groups and geometric analysis: integral geometry, invariant
  differential operators, and spherical functions}, vol.~83.
\newblock American Mathematical Society, 2022.

\bibitem{Susskind:2023rxm}
L.~Susskind, ``{A Paradox and its Resolution Illustrate Principles of de Sitter
  Holography},'' \href{http://arxiv.org/abs/2304.00589}{{\ttfamily
  arXiv:2304.00589 [hep-th]}}.

\bibitem{Harlow:2023hjb}
D.~Harlow and T.~Numasawa, ``{Gauging spacetime inversions in quantum
  gravity},'' \href{http://arxiv.org/abs/2311.09978}{{\ttfamily
  arXiv:2311.09978 [hep-th]}}.

\bibitem{Maldacena:2001kr}
J.~M. Maldacena, ``{Eternal black holes in anti-de Sitter},''
  \href{http://dx.doi.org/10.1088/1126-6708/2003/04/021}{{\em JHEP} {\bfseries
  04} (2003) 021},
\href{http://arxiv.org/abs/hep-th/0106112}{{\ttfamily arXiv:hep-th/0106112
  [hep-th]}}.
%%CITATION = HEP-TH/0106112;%%.

\bibitem{Saad:2019pqd}
P.~Saad, ``{Late Time Correlation Functions, Baby Universes, and ETH in JT
  Gravity},''
\href{http://arxiv.org/abs/1910.10311}{{\ttfamily arXiv:1910.10311 [hep-th]}}.
%%CITATION = ARXIV:1910.10311;%%.

\bibitem{petersen2006riemannian}
P.~Petersen, {\em Riemannian geometry}, vol.~171.
\newblock Springer, 2006.

\bibitem{Chen:2020tes}
Y.~Chen, V.~Gorbenko, and J.~Maldacena, ``{Bra-ket wormholes in gravitationally
  prepared states},'' \href{http://arxiv.org/abs/2007.16091}{{\ttfamily
  arXiv:2007.16091 [hep-th]}}.

\bibitem{Fumagalli:2024msi}
A.~Fumagalli, V.~Gorbenko, and J.~Kames-King, ``{De Sitter Bra-Ket
  Wormholes},'' \href{http://arxiv.org/abs/2408.08351}{{\ttfamily
  arXiv:2408.08351 [hep-th]}}.

\bibitem{Anninos:2020hfj}
D.~Anninos, F.~Denef, Y.~T.~A. Law, and Z.~Sun, ``{Quantum de Sitter horizon
  entropy from quasicanonical bulk, edge, sphere and topological string
  partition functions},'' \href{http://dx.doi.org/10.1007/JHEP01(2022)088}{{\em
  JHEP} {\bfseries 01} (2022) 088},
  \href{http://arxiv.org/abs/2009.12464}{{\ttfamily arXiv:2009.12464
  [hep-th]}}.

\bibitem{Saad:2019lba}
P.~Saad, S.~H. Shenker, and D.~Stanford, ``{JT gravity as a matrix integral},''
\href{http://arxiv.org/abs/1903.11115}{{\ttfamily arXiv:1903.11115 [hep-th]}}.
%%CITATION = ARXIV:1903.11115;%%.

\bibitem{haake1991quantum}
F.~Haake, {\em Quantum signatures of chaos}.
\newblock Springer, 1991.

\bibitem{Stanford:2017thb}
D.~Stanford and E.~Witten, ``{Fermionic Localization of the Schwarzian
  Theory},'' \href{http://dx.doi.org/10.1007/JHEP10(2017)008}{{\em JHEP}
  {\bfseries 10} (2017) 008}, \href{http://arxiv.org/abs/1703.04612}{{\ttfamily
  arXiv:1703.04612 [hep-th]}}.

\bibitem{Kapec:2019ecr}
D.~Kapec, R.~Mahajan, and D.~Stanford, ``{Matrix ensembles with global
  symmetries and \textquoteright{}t Hooft anomalies from 2d gauge theory},''
  \href{http://dx.doi.org/10.1007/JHEP04(2020)186}{{\em JHEP} {\bfseries 04}
  (2020) 186}, \href{http://arxiv.org/abs/1912.12285}{{\ttfamily
  arXiv:1912.12285 [hep-th]}}.

\bibitem{Stanford:2019vob}
D.~Stanford and E.~Witten, ``{JT Gravity and the Ensembles of Random Matrix
  Theory},'' \href{http://arxiv.org/abs/1907.03363}{{\ttfamily arXiv:1907.03363
  [hep-th]}}.

\bibitem{Engelhardt:2023bpv}
N.~Engelhardt, G.~Penington, and A.~Shahbazi-Moghaddam, ``{Twice upon a time:
  timelike-separated quantum extremal surfaces},''
  \href{http://dx.doi.org/10.1007/JHEP01(2024)033}{{\em JHEP} {\bfseries 01}
  (2024) 033}, \href{http://arxiv.org/abs/2308.16226}{{\ttfamily
  arXiv:2308.16226 [hep-th]}}.

\bibitem{Saad:2022kfe}
P.~Saad, D.~Stanford, Z.~Yang, and S.~Yao, ``{A convergent genus expansion for
  the plateau},'' \href{http://dx.doi.org/10.1007/JHEP09(2024)033}{{\em JHEP}
  {\bfseries 09} (2024) 033}, \href{http://arxiv.org/abs/2210.11565}{{\ttfamily
  arXiv:2210.11565 [hep-th]}}.

\bibitem{Aalsma:2020aib}
L.~Aalsma and G.~Shiu, ``{Chaos and complementarity in de Sitter space},''
  \href{http://dx.doi.org/10.1007/JHEP05(2020)152}{{\em JHEP} {\bfseries 05}
  (2020) 152}, \href{http://arxiv.org/abs/2002.01326}{{\ttfamily
  arXiv:2002.01326 [hep-th]}}.

\end{thebibliography}\endgroup

\bibliographystyle{utphys}

\end{document}